\documentclass[final]{siamltex}
\usepackage{epsfig}
\usepackage{graphicx}
\usepackage{amsmath}
\usepackage{amssymb}
\usepackage{setspace}
\usepackage{graphicx}
\graphicspath{{./}{figs/}}
\usepackage{graphicx}

\title{Geometric Registration of High-genus Surfaces}
\author{Chengfeng Wen and Lok Ming Lui}

\begin{document}

\maketitle

\begin{abstract}
This paper presents a method to obtain geometric registrations between high-genus ($g\geq 1$) surfaces. Surface registration between simple surfaces, such as simply-connected open surfaces, has been well studied. However, very few works have been carried out for the registration of high-genus surfaces. The high-genus topology of the surface poses great challenge for surface registration. A possible approach is to partition surfaces into simply-connected patches and registration is done patch by patch. Consistent cuts are required, which are usually difficult to obtain and prone to error. In this work, we propose an effective way to obtain geometric registration between high-genus surfaces without introducing consistent cuts. The key idea is to conformally parameterize the surface into its universal covering space, which is either the Euclidean plane or the hyperbolic disk embedded in $\mathbb{R}^2$. Registration can then be done on the universal covering space by minimizing a shape mismatching energy measuring the geometric dissimilarity between the two surfaces. Our proposed algorithm effectively computes a smooth registration between high-genus surfaces that matches geometric information as much as possible. The algorithm can also be applied to find a smooth and bijective registration minimizing any general energy functionals. Numerical experiments on high-genus surface data show that our proposed method is effective for registering high-genus surfaces with geometric matching. We also applied the method to register anatomical structures for medical imaging, which demonstrates the usefulness of the proposed algorithm.
\end{abstract}

\pagestyle{myheadings}
\thispagestyle{plain}
\markboth{Wen and Lui}{High-genus surface registration}

\section{Introduction}
Registration refers to the process of finding an optimal one-to-one correspondence between images or surfaces. It has been extensively applied to different areas such as medical imaging, computer graphics and computer visions. For example, in medical imaging, registration is always needed for statistical shape analysis, morphometry and processing of signals on brain surfaces (e.g., denoising or filtering). While in computer graphics, surface registration is needed for texture mapping, which aligns each vertex to a position of the texture image, to improve the visualization of the surface mesh. Developing an effective algorithm for registration is therefore very important.

Surface registration between simple surfaces, such as simply-connected open surfaces or genus-0 closed surfaces, has been extensively studied. A lot of effective algorithms have been proposed. However, as far as we know, very few literatures have been reported on the registration between high-genus surfaces. The high-genus topology of the surfaces poses a great challenge to register the surfaces. A possible approach to cope with high-genus surface registration is by introducing cuts to partition the surface into several simply-connected patches. Registration can then be carried out in a patch-by-patch manner. As a result, consistent cuts are required, which is usually difficult to locate and prone to error. Motivated by this, we are interested in developing a geometric registration algorithm for high-genus surfaces, which does not involve the introduction of boundary cuts.

In this paper, we propose an effective way to obtain registrations between high-genus surfaces without introducing any cuts, which matches the geometry as much as possible. The key idea is to conformally parameterize the surface into its universal covering, which is either the 2D Euclidean plane $\mathbb{C}$ or the hyperbolic disk $\mathbb{H}^2$, using the discrete Ricci flow method. Registration can then be done on the universal covering space by minimizing a shape mismatching energy measuring the geometric dissimilarity between the surfaces. Our proposed algorithm effectively computes a smooth registration between high-genus surfaces that matches geometric information as much as possible. To test the performance of the proposed method, numerical experiments have been done on synthetic high-genus surface data. Results show that our proposed algorithm is effective in registering high-genus surfaces with complete geometric matching. The proposed method has also been applied to register anatomical structures for medical imaging, which demonstrates the usefulness of the proposed algorithm.

The rest of the paper is organized as follows. In section 2, we describe some previous works closely related to our paper. In section 3, we introduce some basic mathematical concepts. The proposed algorithm for high-genus surface registration is explained in detail in section 4. The detailed numerical implementation of the algorithm will be described in section 5. In section 6, we show the numerical experimental results. Conclusion and future works are described in section 7.
\section{Previous works}
In this section, we will describe some previous works closely related to our works.

Our proposed algorithm requires surface parameterization of the high-genus surface onto its universal covering space. Surface parameterization has been extensively studied, for which different representations of bijective surface maps have been proposed. Conformal registration, which minimizes angular distortion, has been widely used to obtain a smooth 1-1 correspondence between surfaces \cite{Fischl2, Gu1, Gu3, Haker, Hurdal, Gu2}. For example, Hurdal et al. \cite{Hurdal} proposed to compute the conformal parameterizations using circle packing and applied it to registration of human brains. Gu et al. \cite{Gu1, Gu3, Gu2} proposed to compute the conformal parameterizations of human brain surfaces for registration using harmonic energy minimization and holomorphic 1-forms. Conformal registration is advantageous for it preserves the local geometry well.

Surface registration, which aims to find an optimal one-to-one correspondence between surfaces, has also been extensive studied. Various algorithms have been proposed by different research groups. Landmark-free registration has been proposed to obtain 1-1 correspondences between shapes without feature landmarks. Different algorithms have been proposed to obtain registrations based on the shape information (such as curvatures) defined on the surfaces. Lyttelton et al. \cite{Lyttelton} computed surface parameterizations with surface curvature matching. Fischl et al. \cite{Fischl2} proposed an algorithm for brain registration that better aligns cortical folding patterns, by minimizing the mean squared difference between the convexity of the surface and the average convexity across a set of subjects. Lord et al. \cite{Lord} proposed to match surfaces by minimizing the deviation of the registration from isometry. Yeo et al. \cite{Yeo} proposed the spherical demons method, which adopted the diffeomorphic demons algorithm \cite{Vercauteren}, to drive surfaces into correspondence based on the mean curvature and average convexity. Quasi-conformal mappings have been applied to obtain surface registration with bounded conformality distortion \cite{LuiBHF,LuiBHFHP,LuiQuasiYamabe,LuiBeltramirepresentation}. For example, Lui et al. \cite{LuiBHFHP} proposed to compute quasi-conformal registration between hippocampal surfaces based on the holomorphic Beltrami flow method, which matches geometric quantities (such as curvatures) and minimizes the conformality distortion \cite{LuiBHF}. Most of the above registration algorithms cannot match feature landmarks, such as sulcal landmarks on the human brains, consistently. To alleviate this issue, landmark-matching registration algorithms are proposed by various research groups. Bookstein et al. \cite{Bookstein} proposed to obtain a registration that matches landmarks as much as possible using a thin-plate spline regularization (or biharmonic regularization). Tosun et al. \cite{Tosun} proposed to combine iterative closest point registration, parametric relaxation and inverse stereographic projection to align cortical sulci across brain surfaces. These diffeomorphisms obtained can better match landmark features, although not perfectly. Wang et al. \cite{Wang05,Luilandmark} proposed to compute the optimized conformal parameterizations of brain surfaces by minimizing a compounded energy. All of the above algorithms represent surface maps with their 3D coordinate functions. Special attention is required to ensure the bijectivity of the resulting registration. Besides, smooth vector field has also been proposed to represent surface maps. Lui et al. \cite{Lui10} proposed the use of vector fields to represent surface maps and reconstruct them through integral flow equations. They obtained shape-based landmark matching harmonic maps by looking for the best vector fields minimizing a shape energy. The use of vector fields to represent surface maps makes optimization easier, but they cannot describe all surface maps. Time dependent vector fields can be used to represent the set of all surface maps. For example, Joshi et al. \cite{Joshi} proposed the generation of large deformation diffeomorphisms for landmark point matching, where the registrations are generated as solutions to the transport equation of time dependent vector fields. The time dependent vector fields facilitate the optimization procedure, although it may not be a good representation of surface maps since it requires more memory. Later, Lin et al. \cite{Lin} propose a unified variational approach for registration of gene expression data to neuroanatomical mouse atlas in two dimensions that matches feature landmarks. Again, landmarks cannot be exactly matched. Note that inexact landmark-matching registrations are sometimes beneficial. In the case when landmark points/curves are not entirely accurate, this method is more tolerant of errors in labeling landmarks and gives better parameterization. Most of the above algorithms deal with the registration problem between simply-connected open or closed surfaces.

\section{Mathematical background}
In this section, we describe some basic mathematical concepts related to our algorithms. For details, we refer the readers to \cite{Gardiner}\cite{Lehto}\cite{DiffGeomBook}.

A surface $S$ with a Riemannian metric is called a \emph{Riemann surface}. Given two Riemann surfaces $M$ and $N$, a map $f:M\to N$ is \emph{conformal} if it preserves the surface metric up to a multiplicative factor called the conformal factor. An immediate consequence is that every conformal map preserves angles. With the angle-preserving property, a conformal map effectively preserves the local geometry of the surface structure.

According to the Riemann uniformization theorem, every Riemann surface admits a conformal Riemannian metric of constant Gaussian curvature. Such metric is called the uniformization metric. The uniformization metric for a genus $g=1$ surface induces 0 Gaussian curvature, whereas a genus $g > 1$ surface induces $-1$ Gaussian curvature, which is called the hyperbolic metric of the surface.

Given a high-genus surface $S$ (with genus $g\geq 1$), $S$ is associated with a universal covering space $\widehat{S}\subseteq \mathbb{R}^2$. A universal covering space is a simply-connected space with a continuous surjective conformal map $\pi: \widehat{S}\to S$ satisfying the following: for any $p\in S$, there exists an open neighborhood $U$ of $p$ such that $\pi^{-1}(U)$ is a disjoint union of open sets in $\widehat{S}$. When $g=1$, $\widehat{S}$ is equal to the whole plane $\mathbb{R}^2$. When $g> 1$, $\widehat{S}$ is the unit disk equipped with the hyperbolic metric, which is called the Poincar\`e disk $\mathbb{H}^2$. The Poincar\`e disk $\mathbb{H}^2$ is a unit disk with metric defined as follows:
\begin{equation}
ds^2=\frac{4dzd\bar{z}}{(1-z\bar{z})^2}
\end{equation}

The distance between two points $z$ and $z_0$ on Poincare disk is given by:
\begin{equation}
d(z,z_0) = \tanh^{-1}|\frac{z-z_0}{1-z\bar{z_0}}|
\end{equation}

All rigid motions on Poincar\`e disk are \textit{Mobi\"us transformations}:
\begin{equation}
z\rightarrow e^{i\theta}\frac{z-z_0}{1-z\bar{z_0}}, \quad z_0 \in \mathbb{D},\quad \theta \in [0,2\pi]
\end{equation}

A generalization of conformal maps is the \emph{quasi-conformal} maps, which are orientation preserving homeomorphisms between Riemann surfaces with bounded conformality distortion, in the sense that their first order approximations takes small circles to small ellipses of bounded eccentricity \cite{Gardiner}. Thus, a conformal homeomorphism that maps a small circle to a small circle can also be regarded as quasi-conformal. Surface registrations and parameterizations can be considered as quasi-conformal maps. Mathematically, $f \colon \mathbb{C} \to \mathbb{C}$ is quasi-conformal provided that it satisfies the Beltrami equation:
\begin{equation}\label{beltramieqt}
\frac{\partial f}{\partial \overline{z}} = \mu(z) \frac{\partial f}{\partial z}.
\end{equation}
\noindent for some complex valued Lebesgue measurable $\mu$ satisfying $||\mu||_{\infty}< 1$. $\mu$ is called the \emph{Beltrami coefficient}, which is a measure of non-conformality. In particular, the map $f$ is conformal around a small neighborhood of $p$ when $\mu(p) = 0$. From $\mu(p)$, we can determine the angles of the directions of maximal magnification and shrinking and the amount of them as well. Specifically, the angle of maximal magnification is $\arg(\mu(p))/2$ with magnifying factor $1+|\mu(p)|$; The angle of maximal shrinking is the orthogonal angle $(\arg(\mu(p)) -\pi)/2$ with shrinking factor $1-|\mu(p)|$. The distortion or dilation is given by:
\begin{equation}
K = \frac{1+|\mu(p)|}{1-|\mu(p)|}.
\end{equation}

Thus, the Beltrami coefficient $\mu$ gives us all the information about the properties of the map.

Given a Beltrami coefficient $\mu:\mathbb{C}\to \mathbb{C}$ with $\|\mu\|_\infty < 1$. There is always a quasiconformal mapping from $\mathbb{C}$ onto itself which satisfies the Beltrami equation in the distribution sense \cite{Gardiner}.

\section{Algorithms}
In this section, we explain our algorithm for registering high-genus surfaces in details. The basic idea is to embed the surfaces into their universal covering spaces and register them on the universal covering space. Our proposed algorithm can be divided into three main stages:

\medskip

\begin{enumerate}
\item {\bf Embedding of the high-genus surface into the universal covering space:}  The high-genus surfaces are first conformally parameterized into its universal covering spaces in $\mathbb{R}^2$, which is the Euclidean plane $\mathbb{C}$ for $g=1$ and hyperbolic disk $\mathbb{H}^2$ for $g>1$, using the discrete Ricci flow method.
\item {\bf Computing the initial registration} Harmonic registration between the fundamental domains is computed as an initial registration.
\item {\bf Shape matching registration:} A surface registration which matches the geometry is obtained by minimizing a shape mismatching energy on the universal covering space.
\end{enumerate}

\medskip

In the following, we will describe each stages in detail.

\medskip

\subsection{Embedding of the high-genus surface into the universal covering space}

In this work, the surface registration is computed on the universal covering space in $\mathbb{R}^2$ of the high-genus surfaces. This simplifies the calculation, since all computations can be done in the two dimensional space.

In this work, the embedding of $S$ into its universal covering space $\widehat{S}$ is computed using the Ricci flow method introduced by Gu et al. \cite{JinRicci}\cite{YLYangRicci}. Ricci flow is the process to conformally deform the surface metric $g=(g_{ij}(t))$ according to its induced Gaussian curvature $K(t)$. The process is similar to heat flow on manifolds:
\begin{equation}
\frac{dg_{ij}(t)}{dt}=-2(K(t)-\bar{K})g_{ij}(t)
\end{equation}
where $\bar{K} = 0$ ($g=1$) or $\bar{K} = -1$ ($g>1$) is target curvature. Convergence of this process is guaranteed by Hamilton's theorem. $g(\infty)$ is the desired uniformization metric.

To obtain the embedding, the surface $S$ is firstly sliced along the cut graph $G$ to get the fundamental domain of $S$, denoted as $D$. Let $p\in S$ be a base point on the surface $S$. Then, there are many \textit{closed loops} based at $p$. Two loops $\gamma_1$ and $\gamma_2$ are said to be equivalent if one can be deformed into the other without breaking. Mathematically, there exists a homotopy $H:[0,1]\times [0,1]\to S$ such that $H(0,\cdot) = \gamma_1$ and $H(1,\cdot) = \gamma_2$.  All equivalent closed loops equivalent to each other form an equivalence class. The set of all equivalence classes form a group, which is called {\it fundamental group}, $\pi(S,p)$, of $S$. Suppose $\{a_1,b_1,..., a_i,b_i,..., a_g, b_g\}$ is a basis of $\pi(S,p)$. Slicing along the basis, the high-genus surface will become a simply connected open surface, which is called {\it fundamental domain}.

The fundamental group basis $\{a_1,b_1,a_2,b_2,...,a_g,b_g\}$ is called {\it canonical} if any two loops intersect only at the base point $p$. From algebraic topology, the boundary of the fundamental domain with respect to the canonical loops or cuts is given by
\begin{equation}
a_1b_1a_1^{-1}b_1^{-1}a_2b_2a_2^{-1}b_2^{-1}\cdots a_gb_ga_g^{-1}b_g^{-1}
\end{equation}

In this paper, we apply the greedy approach proposed in \cite{Erickson} to compute the homotopic basis. Each canonical cut is the chosen to be the shortest path in its equivalent class.

\begin{figure*}[t]
\centering
\includegraphics[height=3.5in]{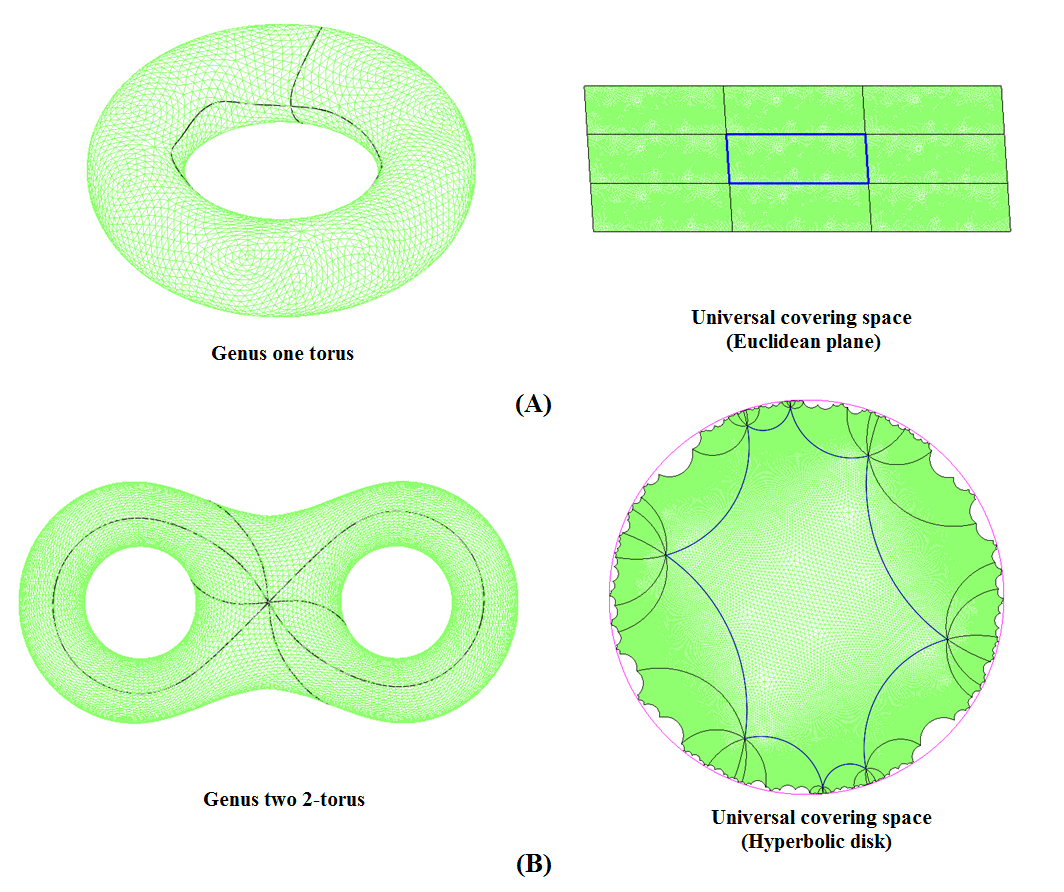}
\caption{(A) shows the genus one torus and its universal covering space (Euclidean plane). (B) shows the genus two 2-torus and its universal covering space (hyperbolic disk). \label{fig:illustrationucs}}
\end{figure*}

With the uniformization metric, the fundamental domain $D$ can be embedded onto a region $\Omega$ in its universal covering space. For genus one closed surface, the universal covering is the Euclidean plane $\mathbb{C}$ (See Figure \ref{fig:illustrationucs}(A)). For genus greater than one, the universal covering is the Poincar\`e disk $\mathbb{H}^2$ (See Figure \ref{fig:illustrationucs}(B)). Let $\pi^{-1}(D)= \bigcup_{\alpha \in U} \widetilde{D}_{\alpha}$, where $U$ is the index set. $\Omega$ belongs to one of the pieces, $\widetilde{D}_{\alpha} \subset \mathbb{H}^2$.  Also, $\widetilde{D}_{\alpha}$ and $\widetilde{D}_{\beta}$ intersect at the boundaries only if $\alpha\neq \beta$. By glueing all $\widetilde{D}_{\alpha}$'s together, the universal covering can be obtained.

Note that the canonical cuts are introduced to obtain the universal covering only. During the registration process, the canonical cuts on the source surface are allowed to move freely on the target surface, since the whole process will be done on the universal coverings. In other words, the correspondences between canonical cuts are not required. It avoids the issue of finding a consistent cuts to obtain the accurate registration.

\subsection{Initial registration between fundamental domains}
We first compute an initial surface registration between two high-genus surfaces $S_1$ and $S_2$ of genus $g$. An initial map can be chosen as the harmonic map by fixing the correspondence of the boundary cuts.

Let $\Omega_1$ and $\Omega_2$ be the canonical fundamental domains of $S_1$ and $S_2$ respectively, computed in the first stage as described in the previous subsection. In this subsection, the metric used is always chosen to be the Euclidean metric if $g=1$ and the hyperbolic metric if $g>1$.

The canonical polygon has $4g$ vertices and hence $4g$ edges. Vertices of the canonical polygon corresponding to the single base point $p$ on the surface. All edges are geodesics.

Let $p_1 \in S_1$ and $p_2\in S_2$ be the base point of $S_1$ and $S_2$ respectively, $p_1$ corresponds to $p_2$. With these base points, $S_1$ and $S_2$ can be conformally mapped to their fundamental domains $\Omega_1$ and $\Omega_2$ in their universal covering spaces (See Figure 1). We denote the conformal parameteriztions by $\phi_1:S_1 \to \Omega_1$ and $\phi_1:S_2 \to \Omega_2$. To obtain an initial registration between $S_1$ and $S_2$, we compute a mapping $g:\Omega_1 \to \Omega_2$ between $\Omega_1$ and $\Omega_2$. $f:=\phi_2^{-1}\circ g \circ\phi_1 : S_1 \to S_2$ gives us an initial mapping between $S_1$ and $S_2$.

Here, we assume corresponding edges between two canonical polygons can be matched. In other words, we assume the boundary condition $h:\partial \Omega_1 \to \partial \Omega_2$ is given, through the arc-length parameterization. Note that the boundary cuts on each surfaces might not exactly correspond to each others. However, since canonical cuts are chosen, the edges corresponds to the shortest loops on the surfaces. As a result, the initial boundary correspondence is a reasonable guess for the initial registration. With the boundary correspondence $h$, a unique harmonic map between the two canonical polygons can be computed.

The harmonic map $g: \Omega_1 \rightarrow \Omega_2$ can be computed by minimizing harmonic energy
\begin{equation}
E(g) = \int_{\Omega_1}{|\nabla g|_{\Omega_2} ^2}, \mathrm{\ \ given\ \ \ } g|_{\partial \Omega_1} = h.
\end{equation}

Minimizing the above energy functional is equivalent to solve following PDE
\begin{equation}
\Delta g = 0 \mathrm{\ \ subject\ to\ \ \ } g|_{\partial \Omega_1} = h.
\end{equation}
where $\Delta$ is Laplace-Beltrami operator under the uniformization metric.



\medskip

This initial registration provide us with a smooth mapping between $S_1$ and $S_2$. Note that there are also other choices of initial maps, such as patch by patch registration or landmark matching registration.

Now, to obtain a geometric matching registration, we propose to refine the registration from the initial registration to match a geometric mismatching energy. The detailed numerical implementation of the initial registration will be described in section \ref{shapematching}.

\subsection{Shape matching registration}\label{shapematching}
In the previous subsection, we use hyperbolic harmonic map between two canonical polygons as initial registration. We assume that the boundary cuts are properly matched. Edges of canonical polygon are (hyperbolic) shortest loops on surface which pass the base point. However, the shortest loops depend on the uniformaization metric, which do not directly take the geometric information of the surfaces into consideration. Constraining the boundary cuts to be exactly matched often induces error in the final registration. To obtain a better geometric matching registration, we propose a variational approach, which minimizes a geometric mismatching energy, without fixing the correspondences of the boundary cuts.

Surface curvatures are important quantities to describe the surface geometry. We therefore consider an energy functional which measures the curvature mismatching under a registration $f:S_1\to S_2$. More specifically, we propose to find an optimal diffeomorphism $f^*:S_1 \to S_2$ which minimizes the following energy functional $E$:
\begin{equation}\label{energyE}
\begin{split}
E(f) = & \frac{1}{2} \int_{S_1}{|\nabla_{g_1} f|_{S_2}^2} +\frac{\alpha^2}{2}\int_{S_1}{(H_1-H_2\circ f)^2} + \frac{\beta^2}{2}\int_{S_1}{(K_1-K_2\circ f)^2}
\end{split}
\end{equation}

\noindent where $H_1,H_2$ are mean curvatures on $S_1$ and $S_2$ respectively, and $K_1,K_2$ are the Gauss curvatures on $S_1$ and $S_2$ respectively.

The first term, which is the harmonic energy, controls the smoothmess of the registration. The last two terms, which measure the mistmatching of surface curvatures, are used to match the surface geometry.

Solving the above variational problem (\ref{energyE}) directly on the surfaces is challenging. To simplify the optimization process, we propose to solve the problem on the universal covering spaces of $S_1$ and $S_2$.

\subsubsection{Optimization on universal covering spaces}
Let $\pi_1: \widetilde{S}_1 \to S_1$ and $\pi_2: \widetilde{S}_2 \to S_2$ be the covering maps of $S_1$ and $S_2$ respectively. Suppose $\pi_1^{-1}(S_1) = \bigcup_{\alpha\in I} \widetilde{D}_{\alpha}^1$, where $I$ is the index set and $\widetilde{D}_{i}^1$ intersects with $\widetilde{D}_{j}^1$ at their boundaries if $i\neq j$. Similarly, we let $\pi_2^{-1}(S_2) = \bigcup_{\beta\in I} \widetilde{D}_{\beta}^2$, where$\widetilde{D}_{i}^2$ intersects with $\widetilde{D}_{j}^2$ at their boundaries if $i\neq j$. We then proceed to look for a diffeomorphism $g^*:\widetilde{S}_1 \to \widetilde{S}_2$, which is the lifting of the optimal registration $f^*:S_1\to S_2$. In other words, we require that
\begin{equation}\label{periodic}
\pi_1^{-1}|_{\widetilde{D}_{\alpha}^1}\circ g^* \circ \pi_2 = f^*, \mathrm{\ for\ any\ } \alpha\in U
\end{equation}

Equation (\ref{periodic}) ensures that $g^*$ satisfies the periodic condition on the covering spaces. In practice, suppose the canonical cuts on $S_1$ is given by $\{a_1,b_1,a_2,b_2,...,a_g,b_g\}$, we require that
\begin{equation}\label{periodic2}
\varphi_i(g^*(a_i)) = g^*(a_i^{-1}) \mathrm{\ and\ } \phi_i(g^*(b_i)) = g^*(b_i^{-1})
\end{equation}
\noindent where $\varphi_i$ and $\phi_j$ are the deck transformations.

Since $g^*$ is the lifting of $f^*$, it minimizes the following energy functional:
\begin{equation}\label{energyEH}
\begin{split}
E_{H}(g) = & \frac{1}{2}\int_{\widetilde{S}_1 }{|\nabla g|^2} + \frac{\alpha^2}{2}\int_{\widetilde{S}_1}{(\tilde{H}_1-\tilde{H}_2\circ g)^2} + \frac{\beta^2}{2}\int_{\widetilde{S}_1}{(\tilde{K}_1-\tilde{K}_2\circ g)^2}
\end{split}
\end{equation}
\noindent subject to the constraint that $\varphi_i(g^*(a_i)) = g^*(a_i^{-1}) \mathrm{\ and\ } \phi_i(g^*(b_i)) = g^*(b_i^{-1})$ for all $1 \leq i\leq g$.

$\{\varphi_1, \phi_1, \varphi_2, \phi_2, ... , \varphi_g, \phi_g \}$ are called the Fuchsian group generators, which are the generators of the Deck transformation group of $S_2$. When $g=1$, $\varphi_i$ and $\phi_i$ are just translations in $\mathbb{R}^2$. When $g>1$, $\varphi_i$ and $\phi_i$ are Mobi\"us transformations of the unit disk, which can be computed explicitly. We will describe the computation of $\varphi_1$. The other Fuchsian group generators can be obtained in the same way. Suppose the starting point and ending points of $a_1$ are $r$ and $s$, and the starting point and ending points of $a_1^{-1}$ are $s'$ and $r'$. We need to look for a Mobi\"us transformation $\varphi_1$ such that $\varphi_1(r) = r'$ and $\varphi_1(s) = s'$. We first compute a Mobi\"us transformation to map $r$ to the origin, which is given by: $\rho_1(z) = (z-r)/(1-\bar{r}z)$. Then, $\rho_1$ maps $\overline{rs}$ to a radial Euclidean line. Let the angle between $\rho_1(\overline{rs})$ and the real axis be $\theta$, and let $\rho_2(z) = e^{-i\theta} z$. Then, $\rho_2\circ\rho_1$ maps $r$ to the origin and $\overline{rs}$ to the real axis. Similarly, we can find Mobi\"us transformation $\rho_1'$ and $\rho_2'$ such that $\rho_2'\circ\rho_1'$ maps $r'$ to the origin and $\overline{r's'}$ to the real axis. The deck transformation $\varphi_1$ is then given by: $\varphi_1 = \rho_1'^{-1}\circ\rho_2'^{-1}\circ \rho_2\circ\rho_1$.

To solve the optimization problem (\ref{energyEH}), we use a splitting method to minimize:
\begin{equation}\label{energysplitting}
\begin{split}
E_{H}(g,h) = & \frac{1}{2}\int_{\widetilde{S}_1}{|\nabla g|^2} + \frac{\mu^2}{2}\int_{\widetilde{S}_1} |g-h|^2 \\
& + \frac{\alpha^2}{2}\int_{\widetilde{S}_1}{(\tilde{H}_1-\tilde{H}_2\circ h)^2} + \frac{\beta^2}{2}\int_{\widetilde{S}_1}{(\tilde{K}_1-\tilde{K}_2\circ h)^2}
\end{split}
\end{equation}

Fixing $g$, we first minimize $E_1(h)$:
\begin{equation}\label{E1}
E_1(h) = \frac{\mu^2}{2}\int_{\widetilde{S}_1} |g-h|^2  + \frac{\alpha^2}{2}\int_{\widetilde{S}_1}{(\tilde{H}_1-\tilde{H}_2\circ h)^2} + \frac{\beta^2}{2}\int_{\widetilde{S}_1}{(\tilde{K}_1-\tilde{K}_2\circ h)^2}
\end{equation}

At each point $p\in \mathbb{H}^2$, we consider the Taylor's expansion of $H_2$ and $K_2$ about $g(p)$,
\begin{equation}\label{Taylor}
\begin{split}
H_2(h)(p)  & \approx H_2(g)(p) + \nabla H_2 (g)(p) \cdot (h-g)(p)\\
K_2(h)(p)  & \approx K_2(g)(p) + \nabla K_2 (g)(p) \cdot (h-g)(p)\\
\end{split}
\end{equation}

Plugging equations (\ref{Taylor}) into equation (\ref{E1}), we look for a small perturbation from $g$ to $h$ such that $E_1$ is minimized. It can be done by solving the following PDE:
\begin{equation}\label{E1PDE}
\mu^2 (g-h) + \alpha^2 (H_1 - H_2(h))\nabla H_2(h) +  \beta^2 (K_1 - K_2(h))\nabla K_2(h) = 0
\end{equation}

In the discrete case, the above problem can be solved by the Guass-Newton method, which will be described in the next section.

Next, fixing $h$, we minimize
\begin{equation}\label{E2}
E_{2}(g) =  \frac{1}{2}\int_{\widetilde{S}_1}{|\nabla g|^2} + \frac{\mu^2}{2}\int_{\widetilde{S}_1} |g-h|^2
\end{equation}

$E_2$ can be minimized by solving the elliptic PDE:
\begin{equation}\label{E2gradient}
\Delta g - \mu^2 (g-h) = 0
\end{equation}

Recall that the registration computed should satisfy the constraint (\ref{periodic}). Hence, we enforce this constraint when solving equation (\ref{E2gradient}). In the discrete case, the above problem becomes a nonlinear system, which can be solved effectively using Newton's method.

In this way, we can minimize $E_H$ alternatively over $g$ and $h$. More specifically, suppose $(g_n, h_n)$ is obtained at the n-th iteration, we fix $g_n$ to obtain $h_{n+1}$ by solving equation (\ref{E1PDE}). We then fix $h_{n+1}$ to obtain $g_{n+1}$ by solving equation(\ref{E2gradient}).

\subsubsection{Preservation of bijectivity}
One crucial issue in computing the surface registration is to preserve its bijectivity. In this work, we propose to enforce the bijectivity using the Beltrami coefficient of the surface map.

Let $g:\widetilde{S}_1 \to \widetilde{S}_2$ be the mapping between the universal coverings of $S_1$ and $S_2$. We need to ensure that $g$ is bijective. Every mapping $g$ is associated with a Beltrami coefficient, $\mu(g)$, which is a complex-valued function defined on $\widetilde{S}_1$. $g$ is bijective if and only if its Jacobian $J_g > 0$ everywhere. Simple checking gives
\begin{equation}
J_g = |\frac{\partial g}{\partial z}|^2 (1-|\mu(g)|^2)
\end{equation}
\noindent Hence, $g$ is bijective if and only if $|\mu(g)| < 1$ everywhere.

Motivated by the above observation, we propose to enforce $\mu(g_n)<1$ in each iterations during the optimization process described in the last subsection. This can be done as follows. Suppose $g_n$ is obtained at the n-th iteration. Let $\epsilon >0$ be a small parameter. We first compute:
\begin{equation}\label{mucrop}
\nu_n = \begin{cases} \max\{|\mu_n|, 1-\epsilon\} \frac{\mu_n}{|\mu_n|}, & \mbox{if } |\nu_n |\neq 0 \\ 0, & \mbox{if } |\nu_n |= 0 \end{cases}
\end{equation}
We then smooth $\nu_n$ by minimizing the following energy functional:
\begin{equation}\label{smoothmu}
\int_{\mathbb{H}^2} |\nabla \nu|^2 +\frac{\lambda }{2}\int_{\mathbb{H}^2} |\nu - \nu_n|^2
\end{equation}

The above minimization problem is equivalent to solving the following PDEs:
\begin{equation}\label{smoothmuPDE}
\Delta \nu + \lambda (\nu - \nu_n)^2 = 0
\end{equation}
\noindent subject to the constraint that for every $1 \leq i \leq g$ $ \nu(x) = \nu(\varphi_i(x))$ for all $x\in a_i$ and $ \nu(y) = \nu(\phi_i(y))$ for all $y\in b_i$.

Once a smooth Beltrami coefficient $\widetilde{\nu}_n$ is obtained, we need to find a quasi-conformal map $f_n$ whose Beltrami coefficient closely resemble to $\widetilde{\nu}_n$. Suppose $f = u+iv$ with Beltrami coefficient $\mu(f) = \rho + i \tau$. We can write $v_x$ and $v_y$ as linear combinations of $u_x$ and $u_y$,
\begin{equation}\label{eqt:linearB1cont}
\begin{split}
-v_y & = \alpha_1 u_x + \alpha_2 u_y;\\
v_x & = \alpha_2 u_x + \alpha_3 u_y.
\end{split}
\end{equation}
\noindent where $\alpha_1 = \frac{(1-\rho)^2 + \tau^2}{1-\rho^2 - \tau^2} $; $\alpha_2 = -\frac{2\tau}{1-\rho^2 - \tau^2} $; $\alpha_3 = \frac{(1+\rho)^2 +\tau^2}{1-\rho^2 - \tau^2} $.

Similarly,
\begin{equation} \label{eqt:linearB2cont}
\begin{split}
-u_y & = \alpha_1 v_x + \alpha_2 v_y;\\
u_x & = \alpha_2 v_x + \alpha_3 v_y.
\end{split}
\end{equation}

Since $\nabla \cdot \left(\begin{array}{c}
-v_y\\
v_x \end{array}\right) = 0$, we obtain
\begin{equation}\label{eqt:BeltramiPDE}
\nabla \cdot \left(D \left(\begin{array}{c}
u_x\\
u_y \end{array}\right) \right) = 0\ \ \mathrm{and}\ \ \nabla \cdot \left(D \left(\begin{array}{c}
v_x\\
v_y \end{array}\right) \right) = 0
\end{equation}

\noindent where $D = \left( \begin{array}{cc}\alpha_1 & \alpha_2\\
\alpha_2 & \alpha_3 \end{array}\right)$.

Therefore, to construct $f_n$, we let $\mu = \nu_n$ and solve equation (\ref{eqt:BeltramiPDE}) subject to the constraint that $\varphi_i(f_n(a_i)) = f_n(a_i^{-1}) \mathrm{\ and\ } \phi_i(f_n(b_i)) = f_n(b_i^{-1})$ for all $1 \leq i\leq g$. The details of the numerical implementation will be explained in the next section.

\medskip

We summarize our proposed high-genus surface registration algorithm as follows.

\medskip

\noindent $\mathbf{Algorithm\ 1:}$ {\it(High-genus surface registration)}\\
\noindent $\mathbf{Input:}$ {\it High-genus surface $S_1$ and $S_2$}\\
\noindent $\mathbf{Output:}$ {\it Geometric matching surface registration $f:S_1\to S_2$}\\
\vspace{-3mm}
\begin{enumerate}
\item {\it  Compute the conformal parameterizations $\phi_1:S_1\to \Omega_1$ and $\phi_2:S_2\to \Omega_2$ of $S_1$ and $S_2$ respectively;}
\item {\it  Compute the initial mapping $f_0$; Let $g_0 = h_0 = f_0$;}
\item {\it Given $(g_n,h_n)$ at n-th iteration, obtain $h_{n+1}$ by fixing $g_n$ and solving equation \ref{E1PDE}; Fixing $h_{n+1}$, obtain $g_{n+1}$ by solving equation \ref{E2gradient};}
\item {\it Compute the Beltrami coefficient $\mu_{n+1}$ of $g_{n+1}$; obtain a smooth Beltrami coefficient $\widetilde{\nu}_{n+1}$ by solving equations \ref{mucrop} and \ref{smoothmuPDE};}
\item {\it Obtain a quasi-conformal map $f_{n+1}$ from  $\widetilde{\nu}_{n+1}$ by solving equation \ref{eqt:BeltramiPDE};}
\item {\it If $||E_H(f_{n+1}) - E_H(f_{n})|| \geq \epsilon$, continue. Otherwise, stop the iteration.}
\end{enumerate}

\section{Numerical implementation}
In this section, we describe in details the numerical implementation of our proposed algorithm. All our computations are done on the universal covering space, which is $\mathbb{C}$ for genus one surfaces and $\mathbb{H}^2$ for high genus surfaces. The universal covering space consists of infinite copies of fundamental domains, which are unique up to deck transformations. Many important operators are identical on each fundamental domains. For example, the Laplace-Beltrami operator, which is crucial in our model, are identical on each fundamental domain, since it is invariant to rigid motions. Based on this observation, the numerical implementation can be done on one piece of the fundamental domain, while allowing its boundary to be mapped freely onto the universal covering space of the target surface. In other words, boundary correspondences between the canonical cuts of the two surfaces are not enforced.

\subsection{Poisson's equation on the universal covering}\label{solvingPoison}

Laplace-Beltrami operator plays a crucial role in our proposed algorithm. Most key steps involve solving a Poisson equation on the universal covering space. In this subsection, we describe how to discretize Laplace-Beltrami operator on the fundamental domain, which can be lifted to the universal covering space. Poisson's equation can then be solved on the universal covering space.\\

On triangle mesh, Laplace-Beltrami operator can be discretized by the cotangent formula:
\[
\Delta_M f(z_i) = \sum_{j\in N_1(i)}{w_{ij}(f(z_j)-f(z_i))}
\]
where $N_1(i)$ is the set of vertex indices of one-ring neighbors of vertex $z_i$; $w_{ij} = \frac{1}{2}(\cot \alpha + \cot \beta)$ where $\alpha$ and $\beta$ are the two angles facing the edge $[v_i,v_j]$. We use $z_i$ denote both the $i^th$ vertex and it's complex coordinate. Then we obtain the Poisson equation in matrix form:
\[
\mathrm{A} f = \mathrm{b};
\]
where $\mathrm{A}$ is a square matrix, $\mathrm{A}(i,j) = w_{ij}, \mathrm{A}(i,i) = -\sum_{j\in N_1(i)}w_{ij}$.

\begin{figure*}[t]
\centering
\includegraphics[height=2in]{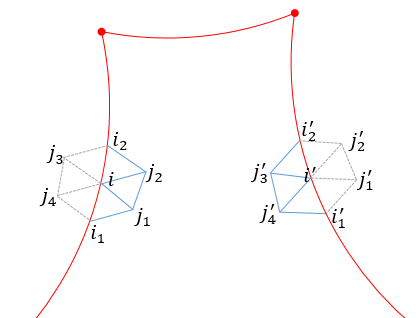}
\caption{fundamental domain on Poincar\`e disk \label{fig:poincare_boundary_illustration}}
\end{figure*}

The computation on each vertex only uses its one-ring neighbor vertices. So we can discretize the Laplace-Beltrami operator on each vertex of the fundamental domain. But on boundary of fundamental domain, the discretization will use vertices outside the fundamental domain (see Figure \ref{fig:poincare_boundary_illustration}).

Every vertex outside the fundamental domain has a unique copy inside the fundamental domain up to a rigid motion. Whenever computation involves vertices outside the fundamental domain, we will refer to its inside copy. Note that this is also valid for base points, so we have a valid discretization on base points. We always fix base points by letting $A(i,i) = 1, A(i,j) = 0$ if $j\neq i$,$\mathrm{b}(i) = f(i)$ for a base point $z_i$. For a vertex on boundary except base points, the discretization will be
\begin{equation}
\Delta_M f(z_i) = \sum_{j\in N_1(i)}{w_{ij}(f(z_j)-f(z_i))}+\sum_{j\in \widetilde{N}_1(i)}{w_{ij}(f(z_{j})-f(z_{i}))}
\end{equation}
where $N_1(i)$ is the set of vertex indices of one-ring neighbors of vertex $z_i$ on the fundamental domain, $\widetilde{N}_1(i)$ is the set of vertex indices of one-ring neighbors of vertex $z_i$ outside the fundamental domain, while $\overset{\circ}{N}_1(i)$ denotes the set of vertex indices of one-ring neighbors of vertex $z_i$ inside the fundamental domain. For example, in figure \ref{fig:Poincare_boundary_illustration}, $N_1(i) = \{i_1,i_2,j_1,j_2\}$, $\widetilde{N}_1(i) = \{j_3,j_4\}$, $\overset{\circ}{N}_1(i) = \{j_1, j_2\}$.

For simplicity, we let $\tilde{z}_i = f(z_i)$. The Laplace-Beltrami operator becomes
\begin{equation}
\Delta_M \tilde{z}_i = \sum_{j\in N_1(i)}{w_{ij}(\tilde{z}_j-\tilde{z}_i)}+\sum_{j\in \widetilde{N}_1(i)}{w_{ij}(\tilde{z}_{j}-\tilde{z}_{i})}
\end{equation}

Suppose $\tilde{z}_j$ is outside the fundamental domain. We denote the the inside copy of vertex $\tilde{z}_j$ by $\tilde{z}_{j'}$. Let $\varphi$ be the deck transformation that moves $\tilde{z}_{j}$ to $\tilde{z}_{j'}$, that is, $\tilde{z}_{j'} = \varphi(\tilde{z}_j)$, we have
\begin{align*}
\Delta_M \tilde{z}_i
& = \sum_{j\in N_1(i)}{w_{ij}(\tilde{z}_j-\tilde{z}_i)}+\sum_{j\in \widetilde{N}_1(i)}{w_{ij}(\tilde{z}_{j}-\tilde{z}_{i})} \\
& = \sum_{j\in N_1(i)}{w_{ij}(\tilde{z}_j-\tilde{z}_i)}+\sum_{j'\in \overset{\circ}{N}_1(i')}{w_{i'j'}(\tilde{z}_{j'}-\tilde{z}_{i'})} \\
& = \sum_{j\in N_1(i)}{w_{ij}(\tilde{z}_j-\tilde{z}_i)}+\sum_{j\in \widetilde{N}_1(i)}{w_{ij}(\varphi(\tilde{z}_j)-\varphi(\tilde{z}_i))}
\end{align*}

The second equality uses the fact that Laplace-Beltrami operator is rigid-motion invariant.


%
With this discretization, the Poison's equation can be rewriten in a matrix form:
\begin{equation}\label{Poisonmatrix}
\mathrm{A}\tilde{z}+\mathrm{Q}(\tilde{z}) = \mathrm{b}
\end{equation}
where $A$ is the matrix representation of the Laplace-Beltrami operator, and $\mathrm{Q}(i,j)$ is a deck transformation that transforms outside neighbor $z_j$ of vertex $z_i$ to its inside copy $z_{j'}$ and is zero elsewhere.

For genus one surfaces, deck transformations are linear translation and so $\mathrm{Q}$ is a linear operator. Combining $\mathrm{Q}$ into $\mathrm{A}$, equation \ref{Poisonmatrix} becomes a linear system and can be solved efficiently.

However, for higher genus surfaces, deck transformations are Mobi\"us transformation, which is nonlinear. Equation \ref{Poisonmatrix} becomes a nonlinear system. It can be solved by Newton's method efficiently. Let $\mathrm{F}(\tilde{z}) =\mathrm{A} \tilde{z}+\mathrm{Q}(\tilde{z})-\mathrm{b}$. $\nabla\mathrm{F} = \mathrm{A+Q'}$ is the gradient of $\mathrm{F}$, where $\mathrm{Q'}$ is computed element-wisely. The problem can then be solved using standard Newton's method:
\begin{enumerate}
\item initialize $\tilde{z}$ by $\tilde{z}_0$ , which is current position;
\item compute $\mathrm{F}(\tilde{z}) = \mathrm{A} \tilde{z}+\mathrm{Q}(\tilde{z})-\mathrm{b}$, if $\|\mathrm{F}(\tilde{z})\|<\epsilon$, stop the process;
\item compute $\nabla \mathrm{F}(\tilde{z}) = \mathrm{A+Q'}(\tilde{z})$, solve $s$ from equation $\nabla \mathrm{F}(\tilde{z}) \cdot s = \mathrm{F}(\tilde{z})$; if $\|s\|<\epsilon$, stop the process; Otherwise, let $\tilde{z} = \tilde{z}-s$ and go to step 2.
\end{enumerate}

The linear equation in step 3 can be solved by LU factorization, which turns out to be quite efficient. In our numerical computation, we observe that the Newton's method converges very quickly: usually two or three iterations will achieve $10^{-10}$ accuracy.


\subsection{Solving energy minimizing problem}
We use an alternating approach to minimize the proposed energy function. In each iteration, we first minimize $E_1(h)$ to get $h$, then minimize $E_2(g)$ to get $g$.

We first discuss the minimization of $E_1(h)$.
With the linear approximation (\ref{Taylor}), we have
\begin{align*}
E_1(h) &= \frac{\mu^2}{2}\int_{\widetilde{S}_1} |g-h|^2  \\
&+ \frac{\alpha^2}{2}\int_{\widetilde{S}_1}{(\tilde{H}_1-\tilde{H}_2(g)-\nabla \tilde{H}_2 (g) \cdot (h-g))^2} \\
&+ \frac{\beta^2}{2}\int_{\widetilde{S}_1}{(\tilde{K}_1-\tilde{K}_2(g)-\nabla \tilde{K}_2 (g) \cdot (h-g))^2}\\
&=\frac{1}{2}\int_{\widetilde{S}_1}
{\left\| \left(
\begin{array}{c}
\alpha (\tilde{H}_1-\tilde{H}_2(g))\\
\beta  (\tilde{K}_1-\tilde{K}_2(g))\\
0
\end{array}\right)
-
\left(
\begin{array}{c}
\alpha\nabla \tilde{H}_2 (g)\\
\beta \nabla \tilde{K}_2 (g)\\
\mu I_{2\times 2}
\end{array}\right)(h-g)
\right\|^2
}
\end{align*}
Then the minimization problem can be solved individually for each vertex $p$ in least square sense:
\begin{align*}
\left(
\begin{array}{c}
\alpha\nabla \tilde{H}_2 (g)\\
\beta \nabla \tilde{K}_2 (g)\\
\mu I_{2\times 2}
\end{array}\right)(h-g)(p)
=\left(
\begin{array}{c}
\alpha (\tilde{H}_1-\tilde{H}_2(g))(p)\\
\beta  (\tilde{K}_1-\tilde{K}_2(g))(p)\\
0
\end{array}\right)
\end{align*}
Let
\[
S=\left(
\begin{array}{c}
\alpha\nabla \tilde{H}_2 (g)\\
\beta \nabla \tilde{K}_2 (g)\\
\mu I_{2\times 2}
\end{array}\right)
, \quad
d = \left(
\begin{array}{c}
\alpha (\tilde{H}_1-\tilde{H}_2(g))\\
\beta  (\tilde{K}_1-\tilde{K}_2(g))\\
0
\end{array}\right)
\]
we have
\[
h(p) = g(p) + (S^TS)^{-1}\cdot ( S^Td)
\]
In computation, the inversion $(S^TS)^{-1}$ can be obtained by Sherman-Morrison formula.
Let $u^T = \frac{\alpha}{\mu}\nabla \tilde{H}_2(g)$, $v^T = \frac{\alpha}{\mu}\nabla \tilde{K}_2(g)$, then $S = \mu(u,v,I)^T$, $S^TS = \mu^2(I+uu^T+vv^T)$. Apply Sherman-Morrison formula twice, we have
\[
(S^TS)^{-1} = \frac{1}{\mu^2}(I-\frac{uu^T+vv^T+(u^T\cdot v^\perp)^2I}{1+u^Tu+v^Tv+(u^T\cdot v^\perp)^2})
\]
Hence we have a simple solution for $h$. If we consider either mean curvature $H$ or Gaussian curvature $K$, i.e., $\beta=0$ or $\alpha=0$, the expression of $h$ can be further simplified. For example, if $\beta = 0$, we have
\[
(S^TS)^{-1} = \frac{1}{\mu^2}(I-\frac{uu^T}{1+u^Tu})
\]
hence,
\[
h(p) = g(p) + \frac{ (\tilde{H}_1-\tilde{H}_2(g))\nabla\tilde{H}_2(g)}{\frac{\mu^2}{\alpha^2}+\nabla \tilde{H}_2(g)^T \nabla \tilde{H}_2(g)}(p)
\]

The minimization of $E_2(g)$ is obtained by solving equation (\ref{E2gradient}):
\begin{equation}
\Delta g - \mu^2 (g-h) = 0
\end{equation}
where Laplace operator is discretized by cotangent formula.\\
Since the operation discussed in section 5.1 will not affect identity matrix, it can be applied to this equation. So we have
\begin{equation}
(\mathrm{A-\mu^2 I}) g +\mathrm{Q}(g)  = -\mu^2 h
\end{equation}
This nonlinear equation is then solved by Newton's method. For genus one surfaces, it is still linear, we can solve it directly.

\subsection{Solving Beltrami equation}
To ensure bijectivity, a smoothing operation on Beltrami coefficient is applied. Then we reconstruct the mapping from smoothed Beltrami coefficient by solving the Beltrami equation.\\
The Beltrami equation is in fact a Poisson equation with a generalized Laplace-Beltrami operator (see equation \ref{eqt:BeltramiPDE}). We can solve the equation as described in section \ref{solvingPoison}. Since we have a generalized Laplace-Beltrami operator, cotangent formula can't be used. We use discretization scheme proposed in \cite{LuiBeltramirepresentation}, which also uses one-ring neighborhood to discretize the generalized Laplace-Beltrami operator. Hence, the method described in section \ref{solvingPoison} can still be applied.

More specifically, the gradient operator $\nabla$ can be discretized by linear approximation. For a triangle $T = (i,j,k)$, $p_i = (x_i,y_i)^T, p_j=(x_j,y_j)^T, p_k = (x_k,y_k)^T$ the coordinates of three vertices, let $\mathrm{e}_i = p_k-p_j, \mathrm{e}_j = p_i-p_k, \mathrm{e}_k = p_j-p_i$, we have
\[
\nabla_T f_i = \frac{1}{4a_T}(f_i \mathrm{t}_i + f_j \mathrm{t}_j+ f_k \mathrm{t}_k)
\]
where $a_T$ is the area of the triangle, $\mathrm{t}_i = \mathrm{e}_i^\perp,\mathrm{t}_j = \mathrm{e}_j^\perp,\mathrm{t}_k = \mathrm{e}_k^\perp$.\\
Then we obtain the discrete gradient operator at vertex $i$:
\[
\nabla f_i = \sum_{T\in N_i}{\frac{1}{4a_T}(f_i \mathrm{t}_i + f_j \mathrm{t}_j+ f_k \mathrm{t}_k)}
\]
where $N_i$ be the collection of neighborhood faces attached to vertex $i$. Note that in the summation we omit the superscripts on $f$ and $t$ to avoid confusion.\\
Similarly, the discretization of divergence operator $\nabla \cdot$ for a vector $F = (u,v)^T$:
\[
\nabla \cdot F_i = \sum_{T\in N_i}\frac{1}{4a_T}(F_i \cdot \mathrm{t}_i + F_j \cdot \mathrm{t}_j+ F_k \cdot \mathrm{t}_k)
\]

The discretization of equation (\ref{eqt:BeltramiPDE}) can be obtained by applying above two formulas.\\
Following the discussion in section \ref{solvingPoison}, the Beltrami equation can be formulated as
\begin{equation}
\mathrm{A}\tilde{z}+\mathrm{Q}(\tilde{z})=\mathrm{b}
\end{equation}

The above equation is linear in the case of genus one surfaces and is nonlinear in the case of higher genus surfaces. By solving the equation, we will get reconstructed quasi-conformal map associated to the smoothed Beltrami coefficient.

\section{Experimental results}
To test the efficacy of the proposed algorithm, experiments have been carried out on synthetic high-genus surface data together with real medical data (vertebrae bone and vestibular system).

\subsection{Synthetic surface data} 
\begin{figure*}[t]
\centering
\includegraphics[height=2.75in]{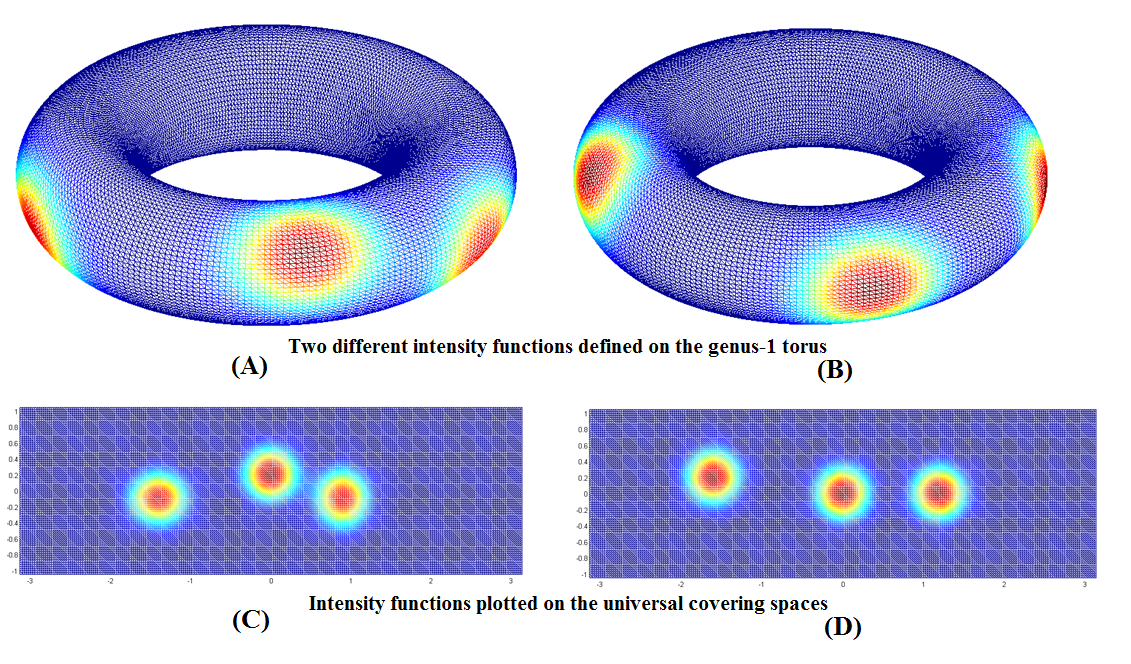}
\caption{(A) and (B) show two genus-1 torus with different intensity functions defined on each of them. (C) and (D) shows the intensity functions plotted on the universal covering spaces of (A) and (B) respectively.\label{fig:tours1}}
\end{figure*}

\begin{figure*}[t]
\centering
\includegraphics[height=2.5in]{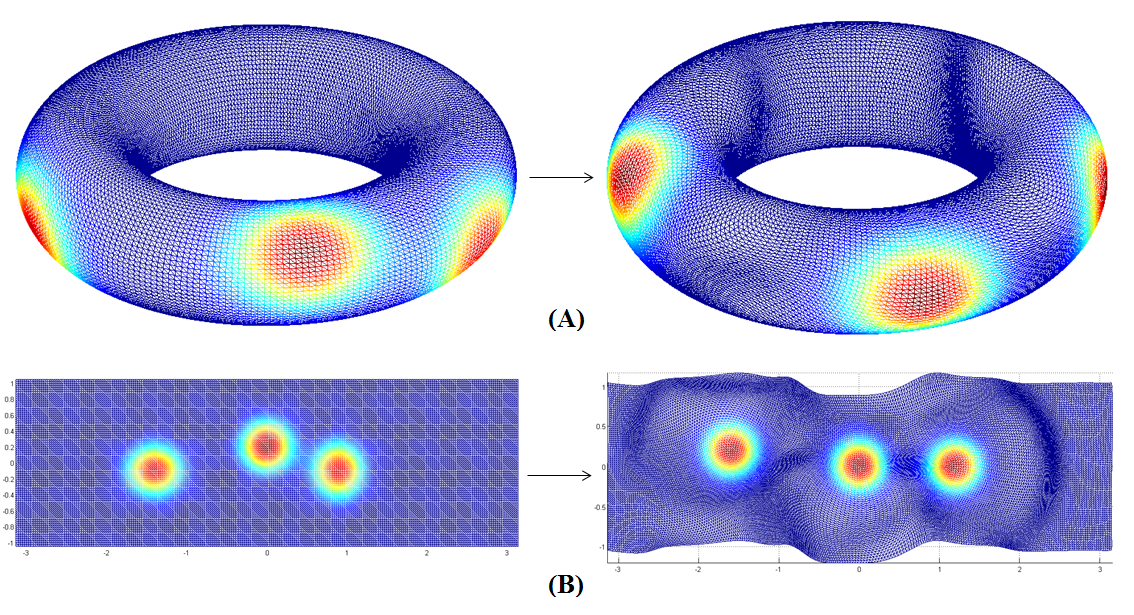}
\caption{(A) shows the registration result that matches the intensity function.  The intensity function defined on $S_1$ is mapped to $S_2$ using the obtained registration. (B) shows the registration result on the universal covering spaces. Note that the boundary cuts are not fixed. They move freely on the universal covering space and satisfy the periodic conditions.\label{fig:torus2}}
\end{figure*}

\begin{figure*}[t]
\centering
\includegraphics[height=1.65in]{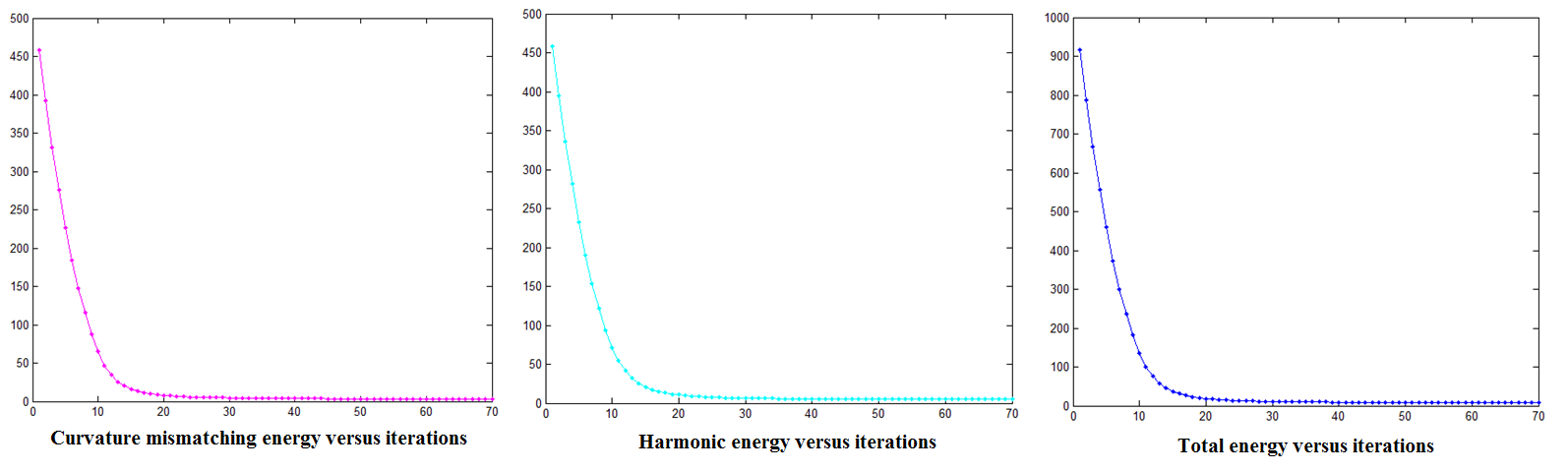}
\caption{The curvature mismatching energy, harmonic energy and total energy versus iterations for the geometric registration problem in Figure \ref{fig:torus2}. \label{fig:torus3}}
\end{figure*}

We first test our algorithm on synthetic surface data.
\paragraph{Example 1} In our first examples, we test the proposed method on a standard torus of genus one. Figure \ref{fig:tours1}(A) and (B) show two genus-1 torus, denoted by $S_1$ and $S_2$ respectively, with different intensity functions defined on each of them. The two surfaces are parameterized onto their universal covering spaces, and registration between the two surfaces is computed on the 2D parameter domains. The intensity functions on each surfaces are plotted on their universal covering spaces, which are shown in (C) and (D). Figure \ref{fig:torus2}(A) shows the registration result that matches the intensity functions. The intensity function defined on $S_1$ is mapped to $S_2$ using the obtained registration. (B) shows the registration result on the universal covering spaces. The intensity functions are perfectly matched under the obtained registration (compared with Figure \ref{fig:tours1}(D)). Note that the boundary cuts are not fixed. They move freely on the universal covering space, which satisfy the periodic conditions. Figure \ref{fig:torus3} shows the curvature mismatching energy, harmonic energy and total energy versus iterations. All of them decrease monotonically as iteration increases. It demonstrates that our algorithm computes the optimized harmonic map between the genus-1 surfaces that matches the intensity functions as much as possible.


\begin{figure*}[t]
\centering
\includegraphics[height=1.55in]{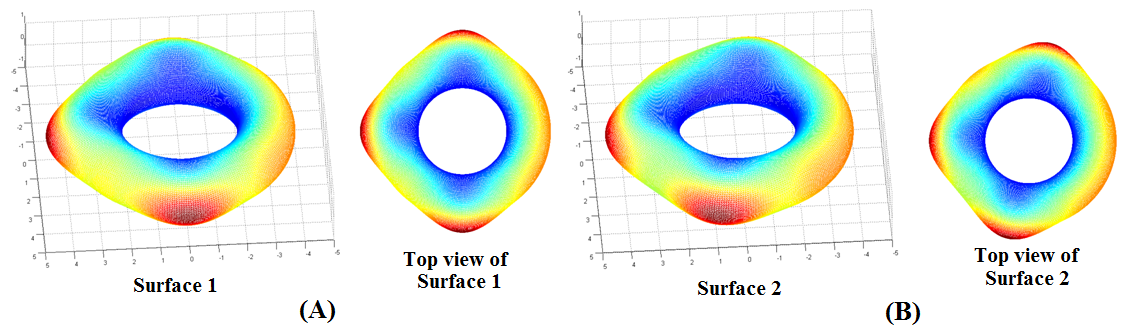}
\caption{ Two synthetic genus-1 surfaces are shown in (A) and (B) respectively. Two bumps are added to each surfaces at different locations. The color-map is given by the mean curvature.\label{fig:torusbump1}}
\end{figure*}

\begin{figure*}[t]
\centering
\includegraphics[height=1.75in]{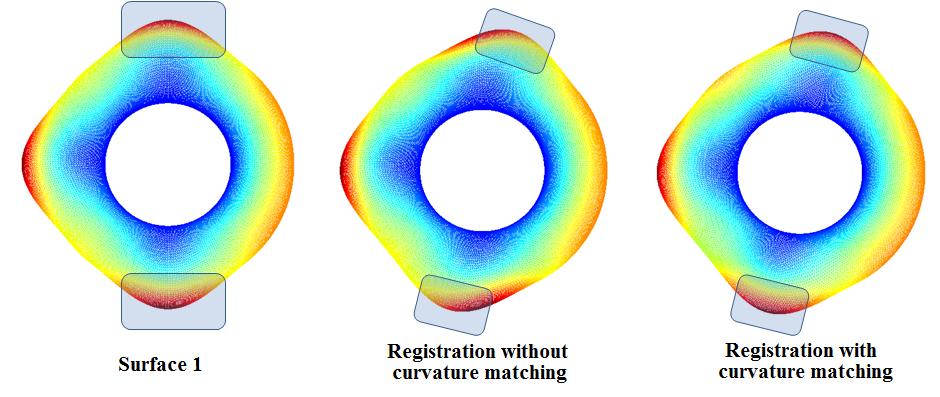}
\caption{ The results of registration without curvature matching and with curvature matching are shown in the figure. The color intensity on surface 1 (given by the mean curvature) are mapped to surface 2 using the obtained registrations. The registration without curvature matching cannot match the feature bumps on the two surfaces, whereas the registration with curvature matching can match the bumps consistently.\label{fig:torusbump2}}
\end{figure*}

\begin{figure*}[t]
\centering
\includegraphics[height=1.65in]{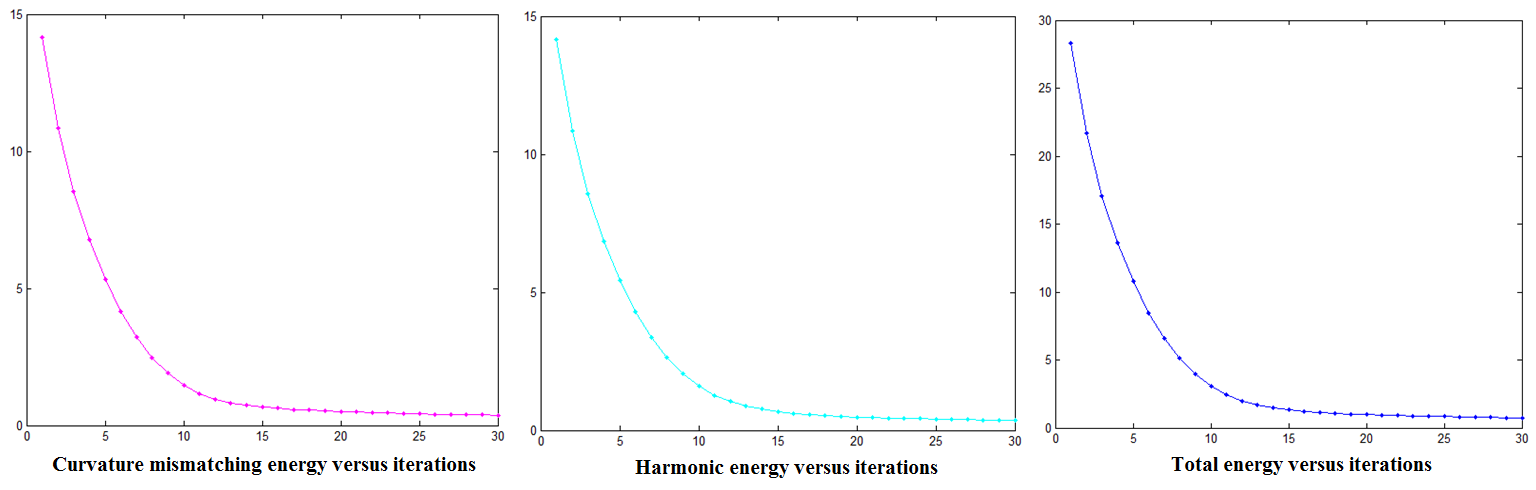}
\caption{ The curvature mismatching energy, harmonic energy and total energy versus iterations for the geometric registration problem in Figure \ref{fig:torusbump2}.\label{fig:torusbump3}}
\end{figure*}

\paragraph{Example 2} We test our proposed algorithm to obtain geometric matching registration between two synthetic genus-1 surfaces through matching their curvatures. Figure \ref{fig:torusbump1}(A) and (B) show two synthetic genus-1 surfaces, with three bumps added to each surfaces located at different positions. The colormaps on each surfaces are given by their mean curvatures. Using our proposed method, we compute both the registration without curvature matching and the registration with curvature matching. The registration result is shown in Figure \ref{fig:torusbump2}. The color intensity on $S_1$ (given by the mean curvature) is mapped to $S_2$ using the obtained registrations. The registration without curvature matching cannot match the feature bumps on the two surfaces, whereas the registration with curvature matching can match the bumps consistently. It illustrates that our proposed method can obtain a better registration that matches geometry between the two surfaces. The curvature mismatching energy, harmonic energy and total energy versus iterations are shown in Figure \ref{fig:torusbump3}. Again, all energies decrease monotonically as iteration increases and converge in about 30 iterations.

\begin{figure*}[t]
\centering
\includegraphics[height=2.75in]{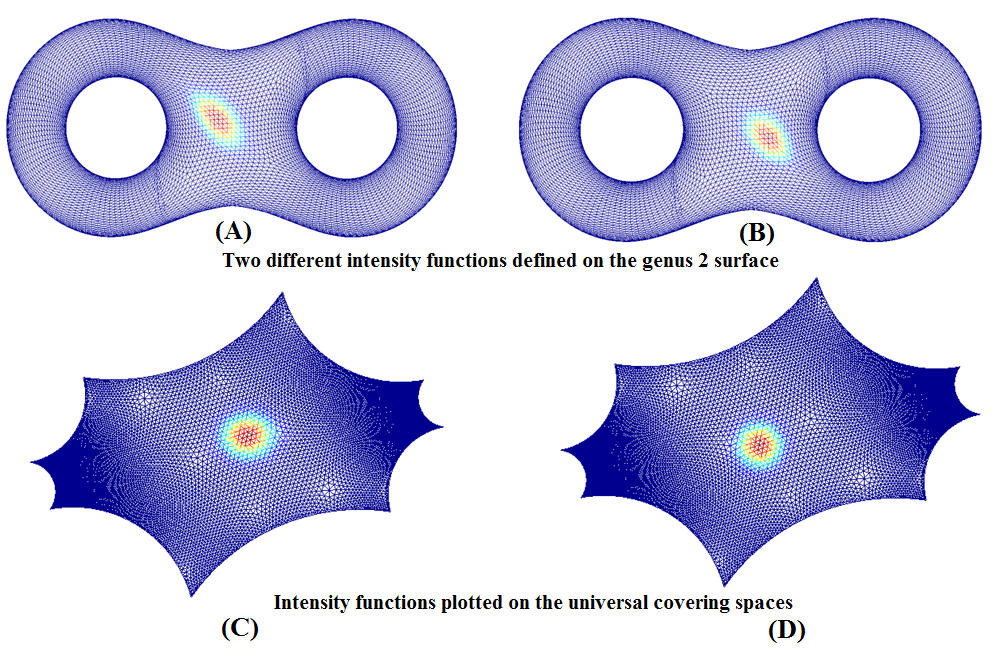}
\caption{(A) and (B) show two genus-2 torus with different intensity functions defined on each of them. (C) and (D) shows the intensity functions plotted on the universal covering spaces of (A) and (B) respectively.\label{fig:genus21}}
\end{figure*}

\begin{figure*}[t]
\centering
\includegraphics[height=2.5in]{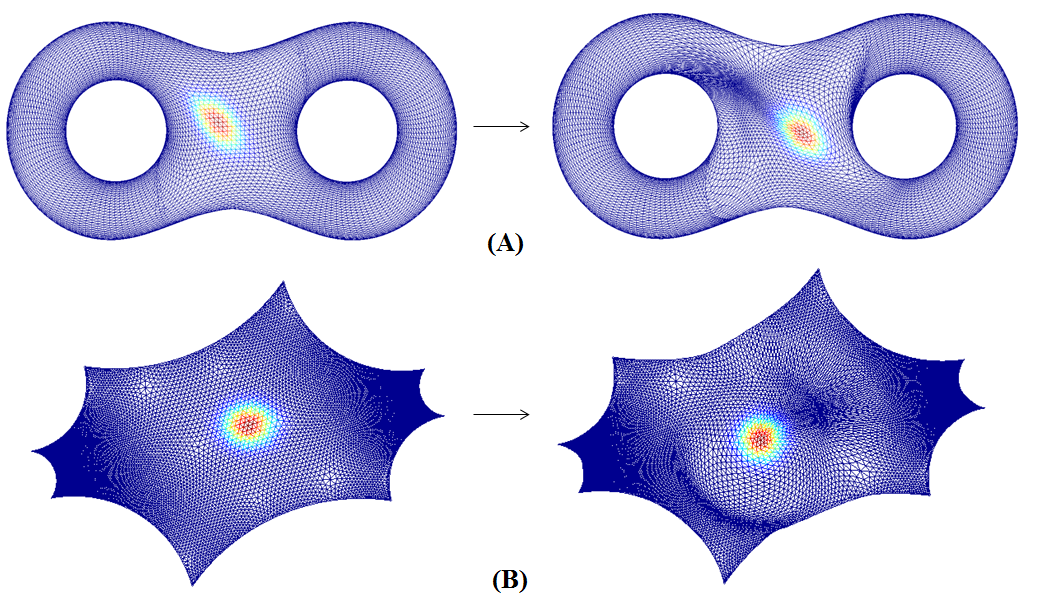}
\caption{(A) shows the registration result that matches the intensity function.  The intensity function defined on $S_1$ is mapped to $S_2$ using the obtained registration. (B) shows the registration result on the universal covering spaces. Note that the boundary cuts are not fixed. They move freely on the universal covering space and satisfy the periodic conditions.\label{fig:genus22}}
\end{figure*}

\begin{figure*}[t]
\centering
\includegraphics[height=1.65in]{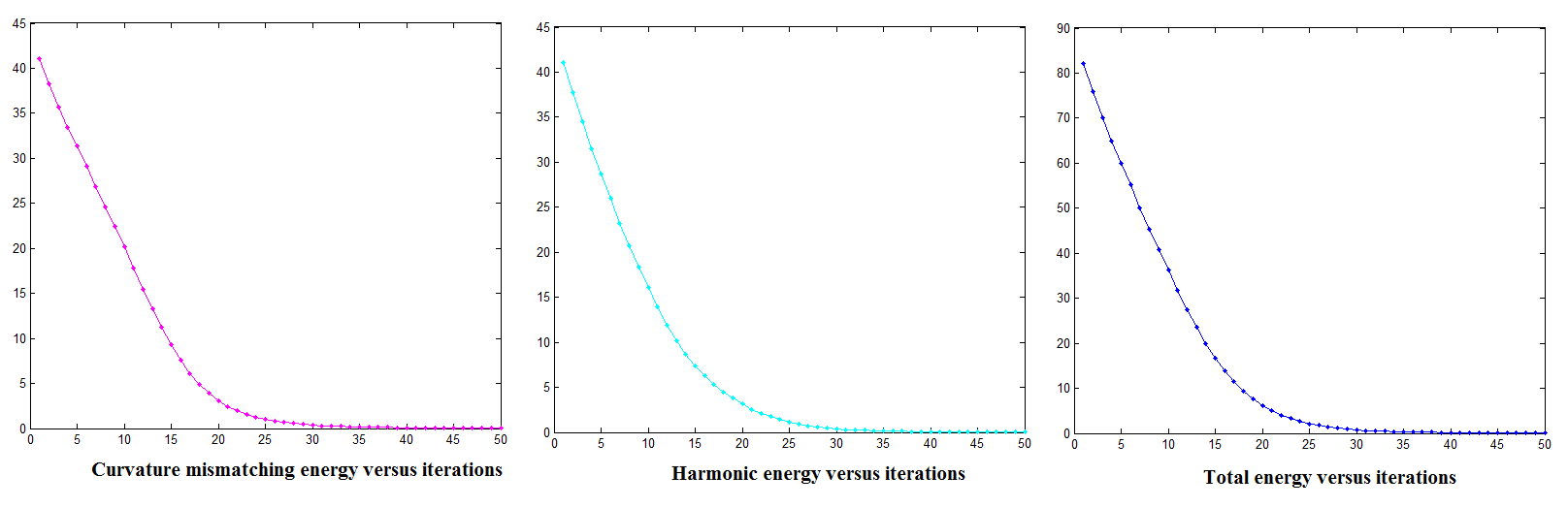}
\caption{The curvature mismatching energy, harmonic energy and total energy versus iterations for the geometric registration problem in Figure \ref{fig:genus22}. \label{fig:genus23}}
\end{figure*}

\paragraph{Example 3} We now test our algorithm on synthetic genus-2 surfaces. Figure \ref{fig:genus21}(A) and (B) show two genus-2 surfaces, denoted by $S_1$ and $S_2$ respectively, with different intensity functions defined on each of them. The two surfaces are parameterized onto their universal covering spaces, and registration between the two surfaces is computed on the 2D parameter domains. The intensity functions on each surfaces are plotted on their universal covering spaces, which are shown in (C) and (D). In Figure \ref{fig:genus22}(A), we shows obtained registration between the two surfaces that matches the intensity functions. The intensity function defined on $S_1$ is mapped to $S_2$ using the obtained registration. (B) shows the registration result on the universal covering spaces. The intensity functions are perfectly matched under the obtained registration (compared with Figure (D)). Again, the boundary cuts are not fixed. They move freely on the universal covering space, which satisfy the periodic conditions. Figure \ref{fig:genus23} shows the curvature mismatching energy, harmonic energy and total energy versus iterations. All of them decrease monotonically as iteration increases. It illustrates that our algorithm computes the optimized (hyperbolic) harmonic map between the genus-2 surfaces that matches the intensity functions as much as possible.

\begin{figure*}[t]
\centering
\includegraphics[height=3.5in]{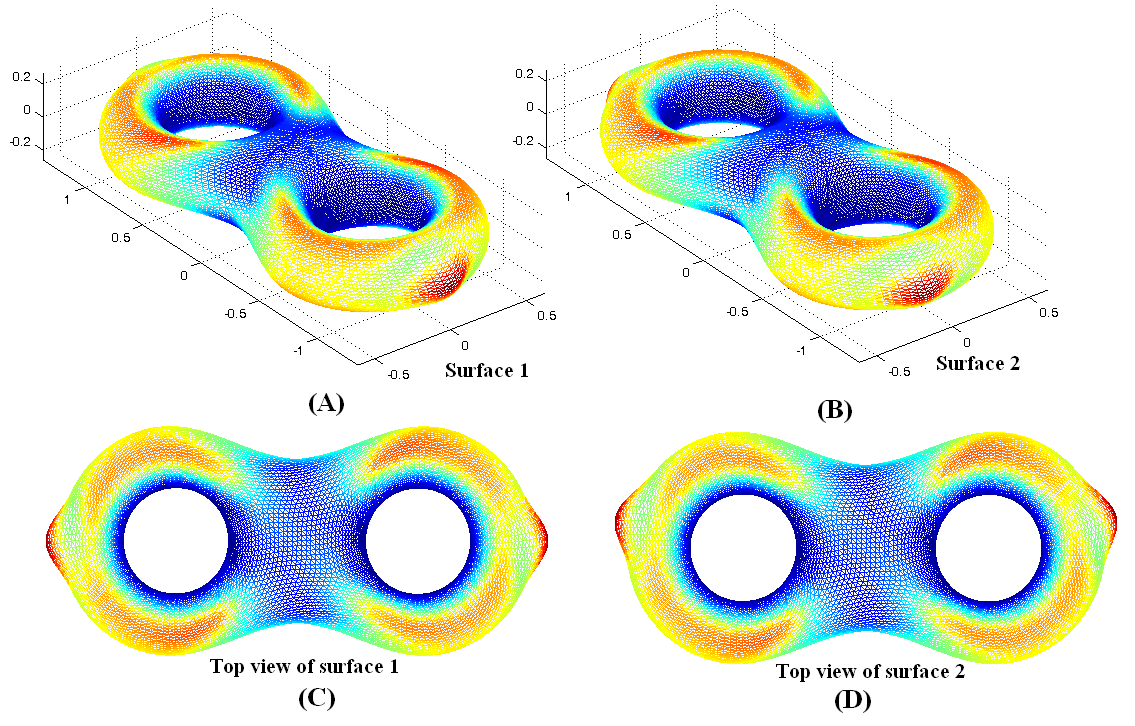}
\caption{Two synthetic genus-2 surfaces are shown in (A) and (B) respectively. Two bumps are added to each surfaces at different locations. The color-map is given by the mean curvature. (C) and (D) shows the top view of the two surfaces. \label{fig:eight1}}
\end{figure*}

\begin{figure*}[t]
\centering
\includegraphics[height=2.75in]{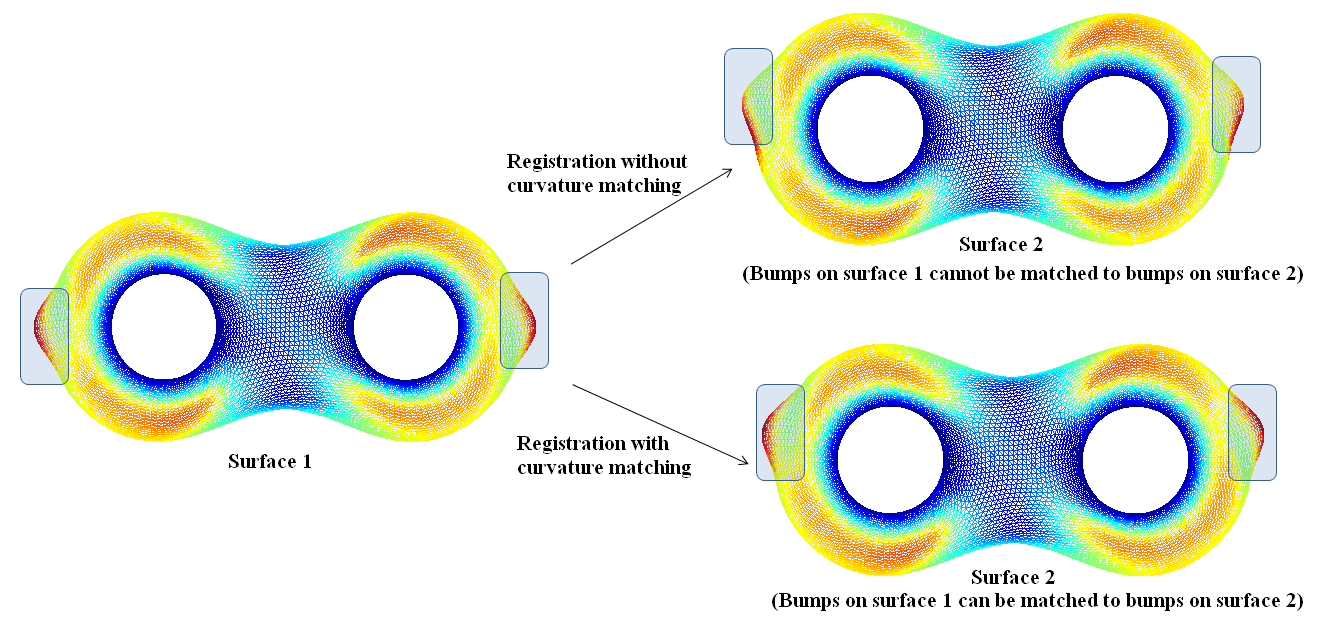}
\caption{The results of registration without curvature matching and with curvature matching are shown in the figure. The color intensity on surface 1 (given by the mean curvature) are mapped to surface 2 using the obtained registrations. The registration without curvature matching cannot match the feature bumps on the two surfaces, whereas the registration with curvature matching can match the bumps consistently.\label{fig:eightregistration}}
\end{figure*}

\begin{figure*}
\centering
\includegraphics[height=1.75in]{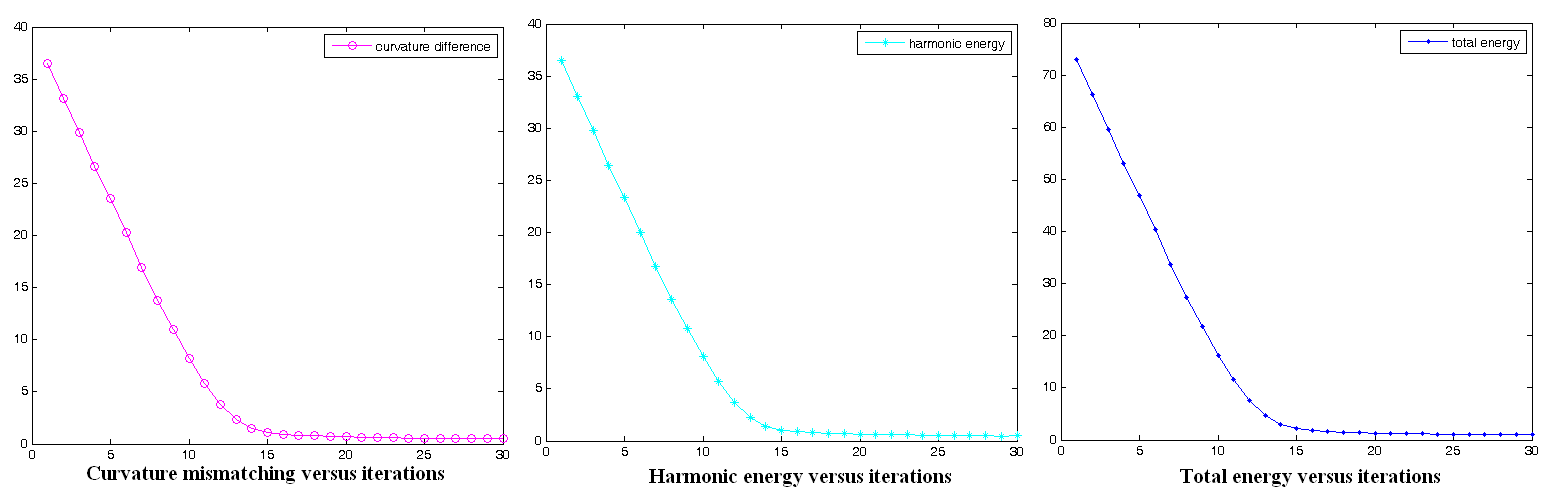}
\caption{The curvature mismatching energy, harmonic energy and total energy versus iterations for the geometric registration problem in Figure \ref{fig:eightregistration}.\label{fig:eightenergy}}
\end{figure*}

\paragraph{Example 4} We also test our method on two synthetic genus-2 surfaces. Figure \ref{fig:eight1}(A) and (B) show two synthetic genus-2 surfaces, with two bumps added to each surfaces located at different positions. The colormaps on each surfaces are given by their mean curvatures. Using our proposed method, we compute both the registration without curvature matching and the registration with curvature matching. The registration results are shown in Figure \ref{fig:eightregistration}. The color intensity on $S_1$ (given by the mean curvature) is mapped to $S_2$ using the obtained registrations. The registration without curvature matching cannot match the feature bumps on the two surfaces (see the regions in the highlighted boxes). It is however observed that the registration with curvature matching can match the bumps consistently. It again demonstrates the effectiveness of our proposed method to obtain a geometric matching registration between genus-2 surfaces. Figure \ref{fig:eightenergy} shows the curvature mismatching energy, harmonic energy and total energy versus iterations. Again, all energies decrease monotonically as iteration increases and converge in about 20 iterations.

\subsection{Real medical data}
In medical imaging, study shape changes of anatomical structures are important for the purpose of disease analysis. To perform shape analysis effectively, an accurate surface registration between anatomical structures is necessary. In this subsection, we will show two applications of our proposed algorithm in medical imaging to register two real medical data, namely, 1. the vertebrae bone and 2. the vestibular system.

\begin{figure*}[t]
\centering
\includegraphics[height=1.85in]{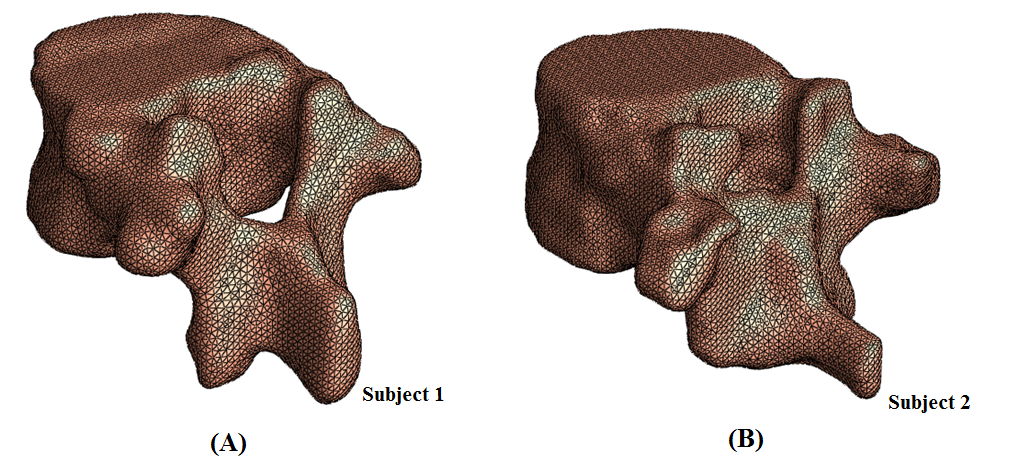}
\caption{The vertebrae bones of genus one of two different subjects. Our goal is to find a geometric matching registration between the two surfaces. \label{fig:boneillustration}}
\end{figure*}

\paragraph{Example 5 (Vertebrae bone)}
The study of morphological changes of the vertebrae is important in detecting vertebral fractures and degenerative shape changes. An accurate and meaningful registration between the vertebrae bone surfaces is therefore important. Using our proposed algorithm, a geometric matching surface registration between different vertebrae bones can be obtained. Figure \ref{fig:boneillustration}(A) and (B) show the vertebrae bones of two different subjects. They are both of genus one. Our goal is to find a geometric matching registration between the two surfaces.

\begin{figure*}[t]
\centering
\includegraphics[height=3.5in]{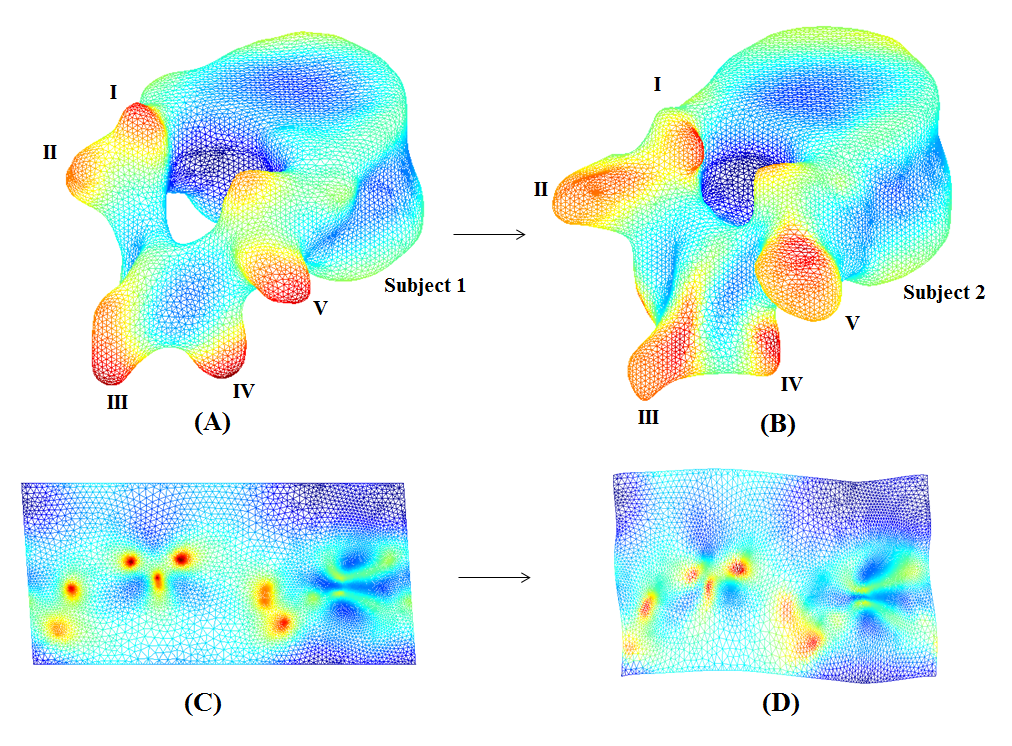}
\caption{The registration result of the vertebrae bones using our proposed algorithm. (A) shows the vertebrae bone surface of subject 1, colored by its mean curvature. The color intensity (given by the mean curvature) on the vertebrae bone of Subject 1 is mapped to the vertebrae bone of Subject 2 in (B), using the obtained registration. Note that the high curvature regions are consistently matched. (C) and (D) shows the registration result on the universal covering spaces. \label{fig:boneregistered}}
\end{figure*}

The registration result of the vertebrae bones using our proposed algorithm is shown in Figure \ref{fig:boneregistered}. (A) shows the vertebrae bone surface of Subject 1, colored by its mean curvature. The color intensity (given by the mean curvature) on the vertebrae bone of Subject 1 is mapped to the vertebrae bone of Subject 2 in (B), using the obtained registration. Note that the high curvature regions are consistently matched. For example, the "hammers" on the vertebae bone of Subject 1 (labeled as region I-VI) are matched consistently with the "hammers" on the vertebae bone of Subject 2. (C) and (D) shows the registration result on the universal covering spaces. Note that the boundary cuts are not fixed. They move freely on the universal covering space, which satisfy the periodic conditions.

\begin{figure*}[t]
\centering
\includegraphics[height=1.5in]{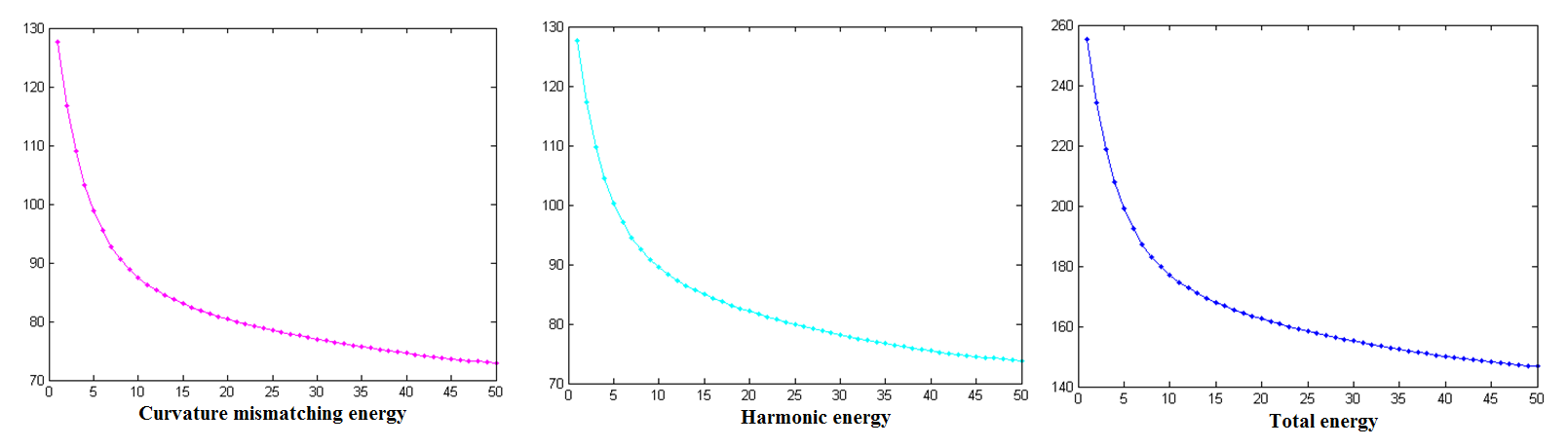}
\caption{The curvature mismatching energy, harmonic energy and total energy versus iterations for the geometric registration problem in Figure \ref{fig:boneregistered}. \label{fig:boneenergy}}
\end{figure*}

Figure \ref{fig:boneenergy} shows the curvature mismatching energy, harmonic energy and total energy versus iterations of our algorithm. All energies monotonically decrease as iteration increases. In particular,  curvature mismatching energy decreases monotonically, which means the optimal map obtained matches curvatures as much as possible.

\paragraph{Example 6 (Vestibular system)}
The vestibular system (VS) is an inner structure of the ear, which is responsible for perception of head movements and sending postural signals to the brain. The shape analysis of the vestibular system plays an important role in understanding a disease called Adolescent Idiopathic Scoliosis (AIS), which is a 3D spinal deformity affecting about 4\% schoolchildren worldwide. It therefore calls for the need to register the vestibular systems. The vestibular system is of genus 3. The high-genus topology of the surface poses great challenges to obtain the surface registration.

\begin{figure*}[t]
\centering
\includegraphics[height=1.85in]{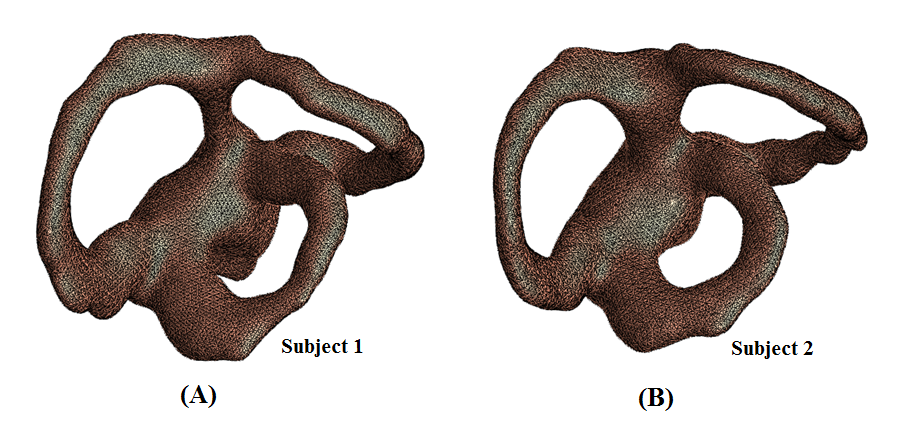}
\caption{The vestibular systems of genus three of two different subjects. Our goal is to find a geometric matching registration between the two surfaces.  \label{fig:vsillustration}}
\end{figure*}

Using our proposed algorithm, we obtain a geometric matching surface registration between the vestibular systems. Figure \ref{fig:vsillustration}(A) and (B) show the vestibular systems of two different subjects. They are both of genus three. Our goal is to find a geometric matching registration between the two surfaces.

\begin{figure*}[t]
\centering
\includegraphics[height=3.75in]{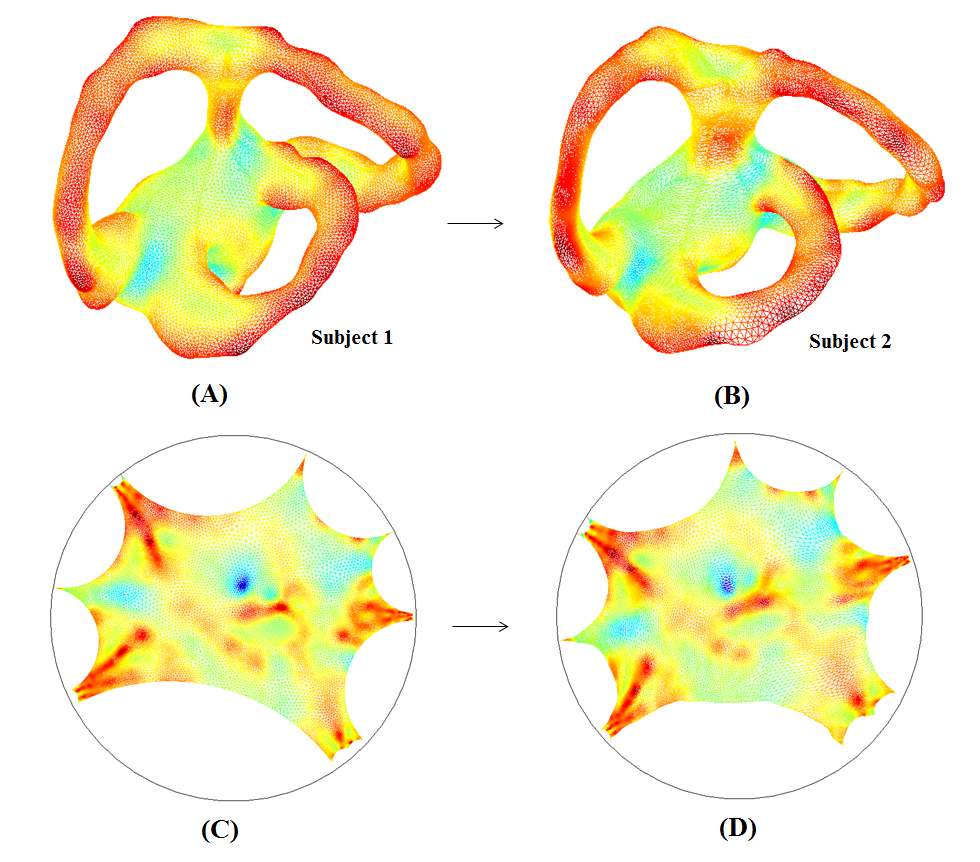}
\caption{The registration result of the vestibular system using our proposed algorithm. (A) shows the vestibular system surface of subject 1, colored by its mean curvature. The color intensity (given by the mean curvature) on the vestibular system of Subject 1 is mapped to the vestibular system of Subject 2 in (B), using the obtained registration. Note that the corresponding regions are consistently matched. (C) and (D) shows the registration result on the universal covering spaces. \label{fig:vsregistered}}
\end{figure*}

The registration result of the vestibular systems using our proposed algorithm is shown in Figure \ref{fig:vsregistered}. (A) shows the vestibular system of Subject 1, colored by its mean curvature. The color intensity (given by the mean curvature) on the vestibular system of Subject 1 is mapped to the vestibular system of Subject 2 in (B), using the obtained registration. Note that the corresponding regions are consistently matched. For example, the three canals of each surfaces are matched consistently. (C) and (D) shows the registration result on the universal covering spaces. Note that the boundary cuts are not fixed. They move freely on the universal covering space, which satisfy the periodic conditions.

\begin{figure*}[t]
\centering
\includegraphics[height=1.5in]{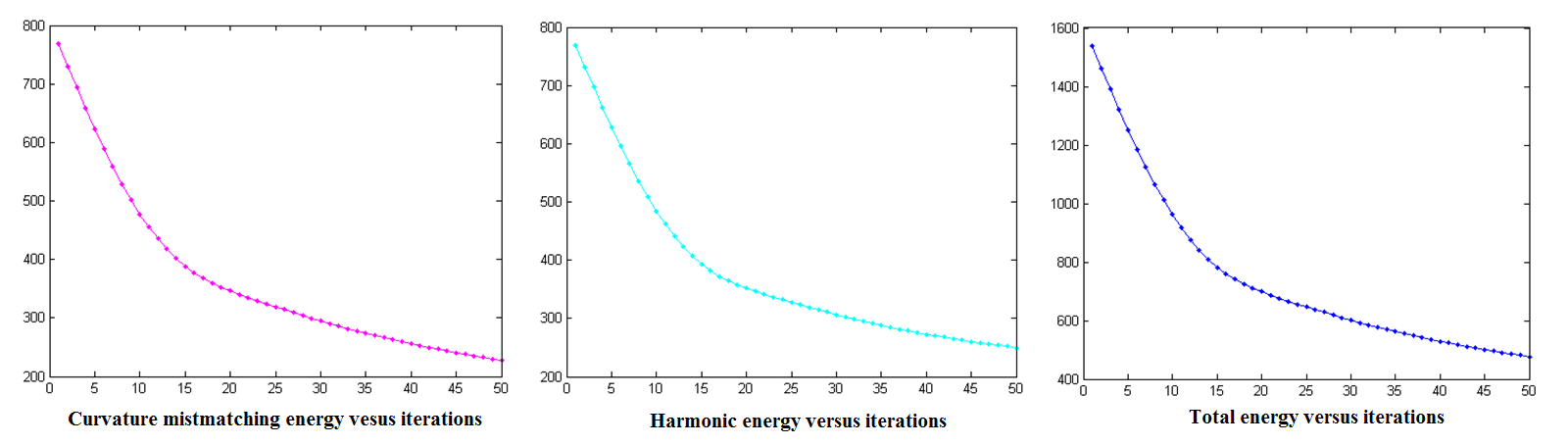}
\caption{The curvature mismatching energy, harmonic energy and total energy versus iterations for the geometric registration problem in Figure \ref{fig:vsenergy}. \label{fig:vsenergy}}
\end{figure*}

Figure \ref{fig:vsenergy} shows the curvature mismatching energy, harmonic energy and total energy versus iterations of our algorithm. All energies monotonically decrease as iteration increases. In particular,  curvature mismatching energy decreases monotonically, which means the optimal registration obtained matches curvatures as much as possible.

\section{Conclusion and future works}
In this work, we propose a method to obtain geometric registrations between high-genus ($g\geq 1$) surfaces, without introducing consistent cuts. The key idea is to conformally parameterize the surface into its universal covering space in $\mathbb{R}^2$. Registration can then be done on the universal covering by minimizing a shape mismatching energy measuring the geometric dissimilarity between the surfaces.  Our proposed algorithm effectively computes a smooth registration between high-genus surfaces that matches geometric information as much as possible. To test the performance of the proposed method, numerical experiments have been done on synthetic high-genus surface data. Results show that our proposed algorithm is effective in registering high-genus surfaces with complete geometric matching. The proposed method has also been applied to registration of anatomical structures for medical imaging, which demonstrates the usefulness of the proposed algorithm. In the future, we will apply the proposed algorithm to register more anatomical structures, such as the vestibular system and the vertebrae bone, for the purpose of disease analysis.

\section*{Acknowledgment} 
Lok Ming Lui is supported by RGC GRF (Project ID: 2130271) and CUHK Direct Grant (Project ID: 2060413). The medical data are provided by CUHK Medical School.

\end{document}